\documentclass[twocolumn,twocolappendix]{aastex7}

\usepackage{amsmath}	

\newcommand{\surfs}{\textsc{surfs}}
\newcommand{\sharkI}{\textsc{shark} v1.1}
\newcommand{\sharkII}{\textsc{shark} v2.0}
\newcommand{\Mhalo}{M_\mathrm{halo}}
\newcommand{\Mbulge}{M_{\star,\mathrm{bulge}}}

\newcommand{\MBH}{M_\bullet}
\newcommand{\fmerge}{f^{}_{M_\bullet,\mathrm{mergers}}}

\newcommand{\DeltaSFMS}{\Delta^{}_\mathrm{SMFS}}
\newcommand{\DeltaMSR}{\Delta^{}_\mathrm{SMSR}}

\newcommand{\DeltaSBHR}{\Delta^{}_\mathrm{SBHMR}}
\newcommand{\DeltaBBHR}{\Delta^{}_{\mathrm{SBHMR}_\mathrm{bulge}}}
\newcommand{\nth}{$^\mathrm{th}$}

\newcommand{\msun}[1]{10^{#1}\ \mathrm{M}_\odot}

\defcitealias{lagos2018}{L18}
\defcitealias{lagos2024}{L24}

\newcommand{\SharkAGNIInp}{Bravo et al. in preparation}

\submitjournal{ApJ}

\begin{document}

\title{The galaxy-AGN scaling relations over 13 billion years in SHARK v2.0 (I): SMBH masses}
\shorttitle{Galaxy-AGN scaling relations in SHARK v2.0 (I)}

\author[0000-0001-5742-7927, gname=Mat\'ias, sname=Bravo]{Mat\'ias Bravo}
\affiliation{Department of Physics \& Astronomy, McMaster University, 1280 Main Street W, Hamilton, ON, L8S 4M1, Canada}
\email[show]{bravosam@mcmaster.ca}
\correspondingauthor{Mat\'ias Bravo}

\author[0000-0003-3021-8564, gname=Claudia, sname=Lagos]{Claudia del P. Lagos}
\affiliation{International Centre for Radio Astronomy Research (ICRAR), M468, University of Western Australia,\\ 35 Stirling Hwy, Crawley, WA 6009, Australia.}
\affiliation{ARC Centre of Excellence for All Sky Astrophysics in 3 Dimensions (ASTRO 3D).}
\affiliation{Cosmic Dawn Center (DAWN), Denmark}
\email{claudia.lagos@uwa.edu.au}

\author[0009-0003-1836-7169, gname=Katy, sname=Proctor]{Katy L. Proctor}
\affiliation{International Centre for Radio Astronomy Research (ICRAR), M468, University of Western Australia,\\ 35 Stirling Hwy, Crawley, WA 6009, Australia.}
\affiliation{ARC Centre of Excellence for All Sky Astrophysics in 3 Dimensions (ASTRO 3D).}
\email{katy.proctor@icrar.org}

\author[0009-0004-1163-0160, gname=\'Angel, sname=Chandro-G\'omez]{\'Angel Chandro-G\'omez}
\affiliation{International Centre for Radio Astronomy Research (ICRAR), M468, University of Western Australia,\\ 35 Stirling Hwy, Crawley, WA 6009, Australia.}
\affiliation{ARC Centre of Excellence for All Sky Astrophysics in 3 Dimensions (ASTRO 3D).}
\email{angel.chandrogomez@research.uwa.edu.au}

\author[0000-0002-4003-0904, gname=Chris, sname=Power]{Chris Power}
\affiliation{International Centre for Radio Astronomy Research (ICRAR), M468, University of Western Australia,\\ 35 Stirling Hwy, Crawley, WA 6009, Australia.}
\affiliation{ARC Centre of Excellence for All Sky Astrophysics in 3 Dimensions (ASTRO 3D).}
\email{chris.power@uwa.edu.au}

\shortauthors{Bravo et al.}

\begin{abstract}
The presence of strong correlations between super-massive black hole (SMBH) masses and galaxy properties like stellar mass have been well-established in the local Universe, but how these scaling relations evolve with cosmic time is yet to be settled in both observations and theoretical models.
Recent works have also highlighted the role of galaxy morphology on the scatter of the SMBH-galaxy mass scaling relations, while the impact of other galaxy properties remains poorly studied, like the role of galaxy environment.
We use the state-of-the-art \sharkII\ semi-analytic model to explore the evolution of these galaxy-SMBH scaling relations to expand the available predictions from theoretical models to contrast with existing and upcoming observations.
We find the relations between SMBH masses and both total and bulge stellar mass predicted by \sharkII\ to be in good overall agreement with observational measurements across a wide range of redshift and stellar masses.
These scaling relations show a significant evolution as a function of cosmic time in \sharkII, with SMBH masses $\sim1$ dex lower at $z=0$ compared to $z=9$ at fixed stellar mass and the scatter increasing by a factor of $\sim2-5$ towards low redshift.
Both relations show a strong dependence with galaxy morphology and the main source for SMBH growth (gas accretion or mergers), with weaker trends with star formation rate, galaxy sizes, and environment.
We find that galaxy morphology alone explains most of the scatter around both scaling relations, with other galaxy properties tying to the SMBH scaling relations through their correlations with morphology.
\end{abstract}


\section{Introduction}

It has become well-established that the masses of super-massive black holes (SMBHs, $\MBH$) correlates with a number of the properties of the host galaxy, like both total and bulge stellar mass ($M_\star$ and $\Mbulge$, respectively), stellar velocity dispersion, and bulge/galaxy sizes \citep[e.g.,][]{magorrian1998,tremaine2002,graham2007,kormendy2013,mcconnell2013}.
These correlations are particularly tight between the SMBH mass and the dispersion-dominated component of a galaxy (bulge in the case of late-type galaxies and whole galaxy in the case of early-types).
The strong connection between SMBH masses and the properties of their host galaxies as been widely interpreted as evidence of the co-evolution of galaxies and their SMBHs, with feedback from active galactic nuclei (AGN) regulating the growth of their stellar mass \citep[e.g.,][]{croton2006,heckman2014,schaye2015}.

There is also an increasing number of works exploring the connection between a number of galaxy host properties and the masses of their SMBHs. 
For example, recent works have found that SMBH masses are the best predictor for the quiescence of central and massive satellite galaxies \citep{mountrichas2023,goubert2024}, and when mass measurements are not available, close proxies for it like the stellar gravitational potential remain the most predictive parameter for galaxy quenching \citep{bluck2023,bluck2024}.
A similar connection has been found between the atomic hydrogen (H\textsc{I}) content of local galaxies and the masses of their SMBHs, with a stronger correlation than between the H\textsc{I} and stellar masses, irrespective of galaxy morpholoy \citep{wang2024}.
More detailed exploration of the relation has also shown that morphology is significantly correlated with SMBH masses \citep[e.g.,][]{vandenbosch2016,dullo2020}, with this correlation being strong enough that early- and late-type galaxies exhibit different SMBH-galaxy scaling relations \citep{graham2023a}.

While the presence of the SMBH-galaxy scaling relations is also clear at higher redshifts \citep[e.g.,][]{jahnke2009,bennert2011,shen2015,dong2016,suh2020}, whether it has evolved with cosmic time or not remains unclear.
Some previous works have argued that observational evidence suggests that the relations evolved with time \citep[e.g.,][]{merloni2010,bennert2011,dong2016,ueda2018,farrah2023}, others that high-redshift measurements are consistent with the low-redshift scaling relations \citep[e.g.,][]{shields2003,shen2015,sun2015,suh2020,li2023b}, and some that not all relations have evolved with time \citep[e.g.,][]{jahnke2009,ding2020}.
More recently, the \textit{James Webb Space Telescope} (\textit{JWST}) has enabled the exploration of the high-redshift $M_\star$-$\MBH$ relation into smaller and more distant galaxies, which suggest that early SMBHs were more massive at fixed stellar mass \citep[e.g.,][]{andika2024,stone2024a}\footnote{Though this proposed evolution may be due to measurement biases \citep[e.g.,][]{ananna2024,lupi2024}.}.


An aspect that has been less explored in the literature is whether environment impacts the evolution of these relations.
Using a small sample of 69 local galaxies, \citet{mcgee2013} found evidence suggesting that satellite and central galaxies following different scaling relations, with satellites having higher SMBH masses than centrals of similar bulge mass below $\Mbulge\sim\msun{11.5}$, but to the authors' knowledge no follow-up has expanded on the results from this work.
More recent observations have found evidence of enhanced accretion into the SMBHs of satellite galaxies undergoing ram-pressure stripping \citep[e.g.,][]{poggianti2017,peluso2022}, which could possibly boost the SMBH growth relative to isolated/central galaxies, but (to the authors' knowledge) this has not been explored.
The study of possible connection between environment and galaxy-SMBH scaling relations seems an area of research that would strongly benefit from new large-scale observations.

Of surveys planned with the \textit{4-metre Multi-Object Spectroscopic Telescope} (\textit{4MOST}), three will target AGN for the measurement of SMBH masses (among other science goals): the \textit{4MOST} AGN Survey \citep{merloni2019}, the Chilean AGN/Galaxy Extragalactic Survey \citep[ChANGES;][]{bauer2023}, and the Time-Domain Extragalactic Survey \citep[TiDES;][]{swann2019}.
These surveys will add $\sim10^6$ galaxies with new SMBH mass measurements across a wide redshift range ($0\lesssim z\lesssim5$), greatly strengthening the statistics for studying the galaxy-SMBH scaling relations.
In paralellel, upcoming redshift surveys like the \textit{4MOST} Hemispheric Survey \citep[4HS;][]{taylor2023} and Wide Area \text{VISTA} Extragalactic Survey \citep[WAVES;][]{driver2019}, and the MOONS Redshift-Intensive Survey Experiment \citep[MOONRISE;][]{maiolino2020} will provide high-quality environmental metrics for millions of galaxies up to $z\sim3$, and the overlap between these and the aforementioned AGN surveys will significantly expand our ability to explore the connection between galaxies, SMBHs, and their environment.
In addition to these surveys, we can expect that continued observations with \textit{JWST} will further increase the number of galaxies with both host and SMBH measurements at more extreme redshifts ($z\gtrsim5$), and also to enable an exploration of their environments. 

The relation between galaxies and their SMBHs has also been widely explored with theoretical models and simulations.
There is no agreement between simulations on whether, and if so, how the normalisation of the $M_\star$-$\MBH$ relation changes with redshift \citep{habouzit2021,habouzit2022}, with some models even predicting a splitting into multiple relations at high redshift \citep{shimizu2024}.
Despite these clear differences between models, there is a overall trend of qualitative agreement with observations. 
For example, several works have found that SMBH masses are a strong predictor of galaxy quenching \citep[e.g.,][]{piotrowska2022,bluck2023}, in line with similar studies from observations.
Detailed simulations have also predicted that environment can play a role in the SMBH feeding and introducing scatter into scaling relations, with satellite galaxies exhibiting more massive SMBHs \citep{shin2012}, mostly driven by tidal stripping and rapid early growth \citep{barber2016,volonteri2016,vanson2019}.

In this work, we use the latest version of the \textsc{shark} semi-analytic model \citep[hereafter L18 and L24, respectively]{lagos2018,lagos2024} to explore the evolution throughout cosmic time of the relation between SMBH masses and both total and bulge stellar masses, exploring how the scatter around both scaling relations relates to a number of galaxy properties and their environment.
We will present the connection between galaxy properties with AGN properties like bolometric luminosity and jet power in a follow-up work (\SharkAGNIInp).
We present an overview of \sharkII\ in Section \ref{S2:SharkV2}, with a focus on the most relevant aspects to the galaxy-BH evolution in the model in Section \ref{S2.1:BHmodel} and our recalibration for the dark matter simulation we use to run \sharkII\ in this work \citep[Planck-Millennium][]{baugh2019} in Section \ref{S2.2:PM}.
We explore the predicted evolution by \sharkII\ of the galaxy-BH scaling relations and how they relate to other galaxy properties in Section \ref{S3:BH}, first in the $\Mbulge$-$\MBH$ plane in Section \ref{S3.1:Mbulge_MBH} followed by the $M_\star$-$\MBH$ plane in Section \ref{S3.2:Mstar_MBH}.
In Section \ref{S4:disc}, we discuss the physical implications derived from the comparison between our predictions to existing observations and how the inform future observations.
Finally, we provide a summary of our work in Section \ref{S5:summary}.

\section{The SHARK v2.0 Semi-Analytic Model}\label{S2:SharkV2}

\sharkII\ is the latest update of the open-source\footnote{Available on GitHub: \url{https://github.com/ICRAR/shark}. Static version used in this work: \citet{sharkv2_b25}.}
and highly modular SAM introduced in \citetalias{lagos2018}.
\sharkI\ has been shown to produce predictions well-matched to a variety of observational measurements: e.g., mass-metallicity relations for gas and stars (\citetalias{lagos2018}), the stellar and gas content scaling relations (\citetalias{lagos2018}, \citealt{hu2020}), the X-ray and radio AGN LFs \citep{amarantidis2019}, the H\textsc{i} content of individual galaxies and galaxy groups/clusters \citep{chauhan2019,chauhan2021}, the scatter around the SFMS \citep{davies2019b,davies2022}, galaxy LFs and number counts from the far-UV to the near-IR and sub-millimetre \citep{lagos2019,lagos2020,stone2024b}, and optical galaxy colours \citep{bravo2020,bravo2022}.

These results have been achieved by including prescriptions for all the physical processes we think shape the formation and evolution of galaxies.
These processes are: 
\begin{enumerate}
    \item collapse and merging of dark matter (DM) haloes;
    \item phase changes of gas between H\textsc{ii}, H\textsc{i} and H$_2$;
    \item accretion of gas onto haloes, which is modulated by the DM accretion rate;
    \item shock heating and radiative metal cooling of gas inside DM haloes, leading to the formation of galactic discs via conservation of specific angular momentum of the cooling gas; \label{item:cooling}
    \item star formation in galaxy discs;
    \item momentum exchange between the gas and stellar components; \label{item:momentum}
    \item stellar feedback from the evolving stellar populations; \label{item:stellar_feedback}
    \item chemical enrichment of stars and gas; \label{item:Z}
    \item seeding of low-mass haloes with SMBHs, and subsequent growth of SMBH via gas accretion and SMBH-SMBH mergers; \label{item:BH}
    \item AGN feedback; \label{item:AGN}
    \item photo-ionisation of the intergalactic medium and intra-halo medium in low mass haloes;
    \item galaxy mergers driven by dynamical friction within common DM haloes, which can trigger starbursts and the formation and/or growth of spheroids; \label{item:mergers}
    \item collapse of globally unstable discs that also lead to starbursts and the formation and/or growth of bulges;
    \item environmental processes affecting the gas content of satellite galaxies. \label{item:environment}
\end{enumerate}

\sharkII\ includes updated physical modelling for several of these processes (\ref{item:momentum}, \ref{item:stellar_feedback}, \ref{item:Z}, \ref{item:BH}, \ref{item:AGN}, \ref{item:mergers}, and  \ref{item:environment}).
We now provide a description of the mechanisms controlling SMBH growth in \sharkII\ (\ref{S2.1:BHmodel}) and of the re-calibration of \sharkII\ for the DM-only simulation we use in this work (\ref{S2.2:PM}), and refer the reader interested in the other improvements in \sharkII\ to \citetalias{lagos2024}.

\subsection{SMBH growth in \sharkII}\label{S2.1:BHmodel}

Galaxies residing in subhaloes with a mass of at least $M_\mathrm{halo,seed}$ are seeded with a SMBH with a mass of $M_\mathrm{BH,seed}$.
Given the similar masses of the DM particles of the DM-only simulations used in \citetalias{lagos2018} and here, we retain the $M_\mathrm{halo,seed}$ of $\msun{10}h^{-1}$, which in practice means that nearly all ($>99.999\%$) of the galaxies with $M_\star\geq\msun{8}$ have been seeded with a SMBH at any given redshift.
The chosen value for $M_\mathrm{BH,seed}$ of $\msun{4}h^{-1}$ has remained unchanged since the first implementation presented \citetalias{lagos2018} to the latest implementation in \citetalias{lagos2024}.
For this work, we tested seed masses as low as $\msun{2}h^{-1}$, finding little difference across cosmic time from our fiducial choice, so we keep the value of $M_\mathrm{BH,seed}$ at $\msun{4}h^{-1}$.

After being seeded, SMBH in \sharkII\ have three growth channels:
\begin{itemize}
    \item SMBH-SMBH mergers when two galaxies merge, which we assume happens instantaneously once the two galaxies merge.
    \item Continuous hot gas accretion from the galaxy halo.
    \item Episodic cold gas accretion after the host galaxy undergoes starburst due to mergers or disc-instabilities.
\end{itemize}
In the new AGN model introduced in \citetalias{lagos2018}, hot gas accretion follows the model introduced by \citet{croton2006} based on a Bondi-Hoyle-like accretion \citep{bondi1952}:
\begin{equation}
    \dot{M}_{\bullet\mathrm{,HG}}=\kappa^{}_\mathrm{AGN}\frac{15}{16}\pi G\mu m^{}_\mathrm{p} \frac{k^{}_\mathrm{B}T_\mathrm{vir}}{\Lambda}M_\bullet, \label{eq:1}
\end{equation}
where $\dot{M}_{\bullet\mathrm{,HG}}$ is the hot gas accretion rate, $\kappa^{}_\mathrm{AGN}$ is a free parameter, $G$ is the gravitational constant, $\mu$ is the mean molecular weight, $m^{}_\mathrm{P}$ is the proton mass, $k^{}_\mathrm{B}$ is Boltzmann's constant, $T_\mathrm{vir}$ is the hot halo gas temperature that depends on the virial mass, $\Lambda$ is a cooling function that depends on both $T_\mathrm{vir}$ and the gas metallicity, and $M_\bullet$ is the SMBH mass.

Cold gas accretion, which only occurs during starbursts, is based on the phenomenological description in \citet{kauffmann2000}.
First, the total cold gas mass accreted in a snapshot is calculated as:
\begin{equation}
    \Delta M_{\bullet\mathrm{,CG}}=f^{}_\mathrm{SMBH}\frac{M_\mathrm{CG}}{1+\left(v^{}_\mathrm{SMBH}/v^{}_\mathrm{vir}\right)}, \label{eq:2}
\end{equation}
where $f^{}_\mathrm{SMBH}$ and $v^{}_\mathrm{SMBH}$ are free parameters controlling the normalisation of the $\Mbulge$-$\MBH$ relation, $M_\mathrm{CG}$ is the cold gas mass of the galaxy during the starburst, and $v^{}_\mathrm{vir}$ is the virial veolocity.
We then use $\Delta M_{\bullet\mathrm{,CG}}$ to calculate the cold gas accretion rate as: 
\begin{equation}
    \dot{M}_{\bullet\mathrm{,CG}}=\frac{\Delta M_{\bullet\mathrm{,CG}}}{\tau^{}_{\bullet\mathrm{,CG}}}=\Delta M_{\bullet\mathrm{,CG}}\left(\frac{e^{}_\mathrm{SB}r^{}_\mathrm{bulge}}{v^{}_\mathrm{bulge}}\right)^{-1}, \label{eq:3}
\end{equation}
where $\tau^{}_{\bullet\mathrm{,CG}}$ is the accretion timescale (assumed to be of the order of the bulge dynamical timescale), $e^{}_\mathrm{SB}$ is a e-folding free parameter, and $r^{}_\mathrm{bulge}$ and $v^{}_\mathrm{bulge}$ are the radius and circular velocity of the gas in the bulge, respectively.

While we do not explore it further in this work, with \sharkII\ we also include three choices for modelling SMBH spin, for which we provide a brief overview here (we refer the interested reader to section 3.3 and appendix B of \citetalias{lagos2024} for further details).
We define the dimensionless BH spin vector as:
\begin{equation}
    \textbf{a} \equiv \frac{\textbf{J}_\bullet}{J_\mathrm{max}} = \frac{c } {G M^2_\bullet}\textbf{J}_\bullet,
\end{equation}
where $c$ is the speed of light, and $\textbf{J}_\mathrm{BH}$ is the angular momentum of the SMBH.
The three SMBH spin models in \sharkII\, which provide values for the norm of the dimensionless BH spin vector, $a = |\textbf{a}|$, are the following:
\begin{itemize}
    \item \texttt{constant}.
    This model makes the simple assumption that SMBH spin is constant (i.e., does not evolve with time) and that the spin of all SMBHs is the same.
    The value of the spin is calculated based on the assumed radiation efficiency of the SMBH \citep[following][]{bardeen1972}, set by the free parameter \texttt{nu\_smbh}, which in \sharkII\ has a default value of 0.1 (equivalent to $a=0.67$).
    \item \texttt{volonteri07}.
    This model follows a scaling relation based on the $\MBH$-$a$ relation for a warped accretion disc described by \citet{volonteri2007}, where they found that higher $\MBH$ correlates with higher $a$, given by:
    \begin{equation*}
        a = 0.305 \log_{10}(\MBH/M_\odot) - 1.7475,
    \end{equation*}
    where we limit the values of the SMBH spin to the range $a\in[0,1]$, i.e, we only consider SMBHs to be co-rotating.
    \item \texttt{griffin19}.
    This model implements the more complex BH spin evolution model presented in \citet{griffin2019}, which was itself an updated version of the model in \citet{fanidakis2011}, that describes the change in SMBH due to gas accretion and SMBH mergers.
    This model includes three sub-models describing gas accretion into the SMBH (prolonged accretion, self-gravitating accretion disc, and warped accrection disc), all which are implemented in \sharkII\ and set by the \texttt{accretion\_disk\_model} parameter, with the default being that for a warped accretion disc (\texttt{accretion\_disk\_model = warpeddisk}).
    In this model, we allow the SMBH spin to take values in the range $a \in[-1, 1]$, i.e., to become counter-rotating.
\end{itemize}

\subsection{\sharkII\ on the Planck-Millennium simulation}\label{S2.2:PM}

\begin{deluxetable}{lCC}
    \tablehead{
        \colhead{Parameters} & \colhead{L210N1536} & \colhead{PM}\\
    }
    \tablecaption{
        Parameters of medi-SURFS L210N1536 and Planck-Millennium simulations.\label{tab:simulations}
    }
    \startdata
        Box size [cMpc $h^{-1}$] & 210 & 542.16 \\
        Number of particles & 1536^3 & 5040^3\\
        Particle Mass [M$_{\odot}$ $h^{-1}$] & 2.21\times 10^8 & 1.06\times 10^8\\
    \enddata
\end{deluxetable}

In \citetalias{lagos2024}, we used the L210N1536 simulation of the Synthetic UniveRses For Surveys suite of DM-only simulations \citep[\surfs;][]{elahi2018a}.
In order to explore the most massive SMBH and brightest AGN, for this work we re-tuned \sharkII\ to use the much larger Planck-Millennium $N$-body simulation \citep[PM hereafter;][]{baugh2019}.
The cosmological parameters correspond to a total matter, baryon and $\Lambda$ densities of $\Omega_\mathrm{m}=0.307$, $\Omega_\mathrm{b}=0.04825$, and $\Omega_\mathrm{\Lambda}=0.693$, respectively, with a Hubble parameter of $H_0=100h$ Mpc km s$^{-1}$ with $h=0.6777$, scalar spectral index of $n_\mathrm{s}=0.9611$ and a power spectrum normalization of $\sigma_8=0.8288$, which we note are close to the values adopted in for \surfs.
Technical specifications of the simulation are presented in Table \ref{tab:simulations}.
PM has $271$ snapshots containing the full particle and halo data.
Given the similar mass resolutions of L210N1536 and PM, we adopt the same stellar mass limit employed in previous works (e.g., \citetalias{lagos2018}, \citetalias{lagos2024}) using both \sharkI\ and \sharkII\ of $M_\star>\msun{8}$, leading to a sample size of $\sim1.7\times10^7$ ($\sim2.3\times10^5$) galaxies at $z=0$ ($z=9$).

The halo catalogues for PM were constructed using \textsc{subfind} \citep{springel2001} and the merger trees were built using \textsc{d-trees}+\textsc{d-halo} \citep{jiang2014}.
We use the latest addition to \sharkII\ presented in \citet{chandro2025} , which implements fixes for numerical artefacts affecting merger tree data (mass-swapping events between central and satellite subhaloes and the sudden emergence of massive transient subhaloes at late cosmic times). 
We note that their analysis finds that the PM merger trees are also significantly affected by these issues, though \citet{chandro2025} also demonstrated that these issues can be effectively minimised at the semi-analytic model level, which is what we do here.
Halos with $\geq15$ particles are included in the catalogues, giving a minimum halo mass of $1.59\times\msun{9}h^{-1}$.
Note that this is different to the adopted halo catalogues and merger trees in \citetalias{lagos2024}, in which we process L210N1536 simulation of the Synthetic UniveRses For Surveys suite of DM-only simulations \citep[\surfs;][]{elahi2018a} using \textsc{VELOCIraptor} \citep{poulton2018,canas2019,elahi2019a} for the halo catalogues and \textsc{TreeFrog} \citep{elahi2019b} for the merger trees.

\begin{deluxetable}{LCCc}
    \tablehead{
        \colhead{Parameter} & \colhead{Value in \citetalias{lagos2024}} & \colhead{Value in this work} & \colhead{Calibration}\\
    }
    \tablecaption{
        Re-tuned parameters in this work (left column), values adopted in \citet{lagos2024} for L210N1536 (second column), values we adopt in this work for PM (third column), and the calibration method used (fourth column).\label{tab:sharktuned}
    }
    \decimals
    \startdata
         f^{}_\mathrm{SMBH} & 0.01 & 0.007 & Manual \\
         e^{}_\mathrm{SB} & 15.0 & 2.0 & Manual\\
         \kappa & 10.307 & 10.307 & \texttt{optim} \\
         \kappa^{}_\mathrm{jet} & 0.02285 &  0.02285 & \texttt{optim} \\
         \Gamma_\mathrm{thresh} & 10.0 & 10.0 & \texttt{optim} \\
         \tau^{}_\mathrm{reinc} & 21.53 & 17.384 & \texttt{optim} \\
         \gamma & -2.339 & -2.3535 & \texttt{optim} \\
         M_\mathrm{norm} & 1.383\times10^{10} & 4.6159\times10^{10} & \texttt{optim} \\
    \enddata
\end{deluxetable}

\citet{gomez2022} analysed the effect of using \textsc{VELOCIraptor}+\textsc{d-trees} and \textsc{subfind}+\textsc{d-trees} in the \textsc{galform} semi-analytic model \citep{lacey2016,baugh2019} and concluded that the effect of using these different combinations of halo finders and merger tree builders had no discernible consequence on the stellar masses, SFRs, and gas masses (both cold and hot).
However, the number of satellite/central galaxies depends slightly on the combination used, with \textsc{subfind}+\textsc{d-trees} producing less satellite subhaloes than \textsc{VELOCIraptor}+\textsc{d-trees} by a factor of $\sim2$ (and hence less satellite galaxies).
From testing \sharkII\ with \surfs\ and PM we reached the same conclusions, though we leave a detailed comparison for future work.

\begin{figure*}
    \centering
    \includegraphics[width=\linewidth]{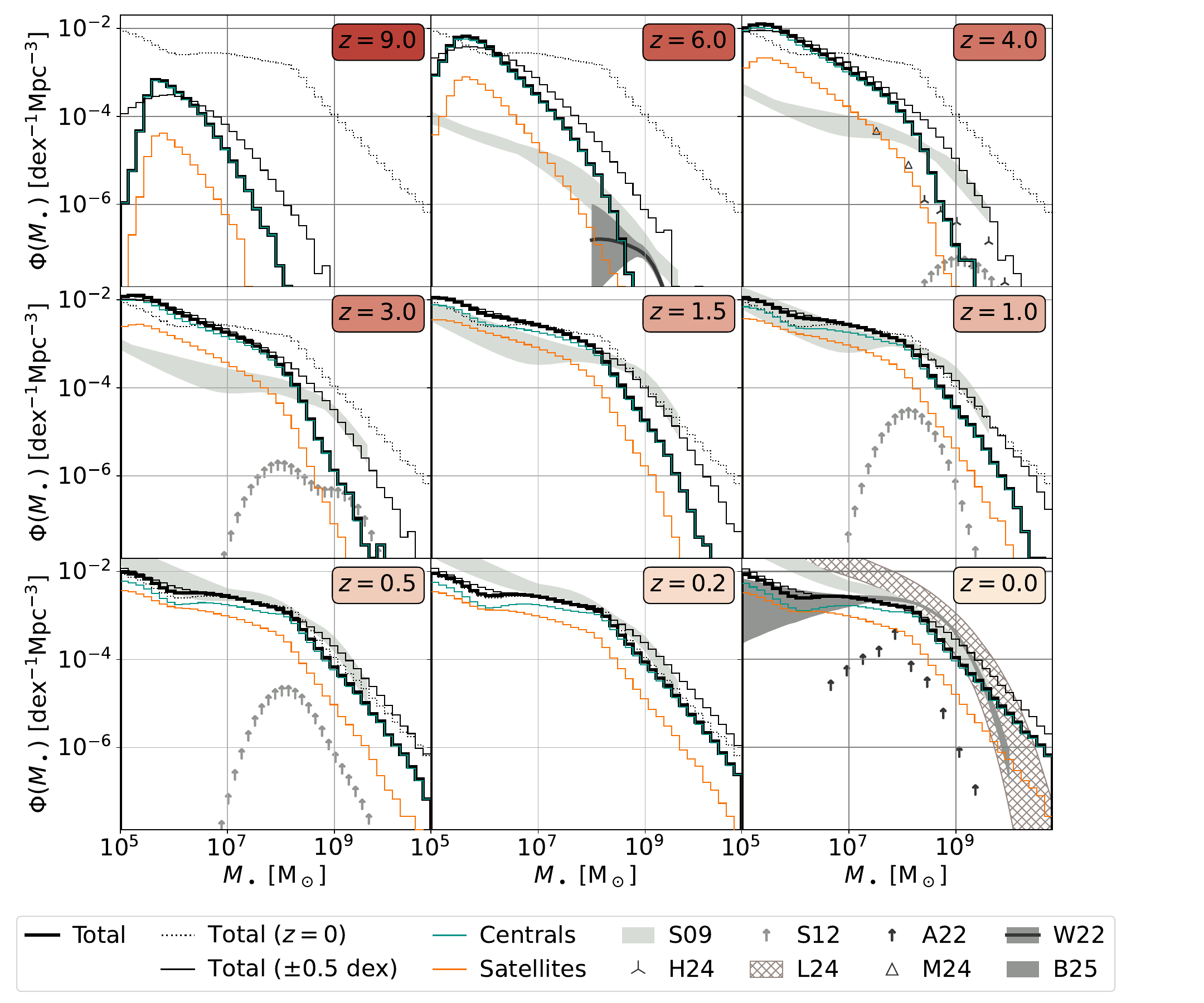}
    \caption{Comparison of the BHMF predicted by \sharkII\ from $z=9$ to $z=0$ to observational results from the literature.
    Each panel corresponds to the SMBH mass function at the redshift indicated at the top right corner.
    The lines show the results from \sharkII, the total BHMF with the thick black line, the total BHMF with a 0.5 dex scatter applied to $\MBH$ with the thin (solid) black line, and the contribution of central and satellites in teal and orange respectively.
    The $z=0$ BHMF is repeated in the other panels with the dotted black line to highlight the time evolution of the BHMF.
    Observational data shown: \citet{shankar2009} with grey bands, \citet{wu2022b} with a solid dark grey line and shaded area at $z=6$, single Schechter fit from \citet{liepold2024} with a hatched reddish grey at $z=0$, \citet{burke2025} with solid dark grey line and shaded area at $z=0$, the lower limits from \citet{shen2012} with light grey arrows, and the lower limits from \citet{ananna2022,he2024,matthee2024} with dark grey arrows, three-pointed stars, and triangles, respectively.}
    \label{fig:BHMF}
\end{figure*}

The combination of the differences in both cosmology and halo finding lead to (mild) differences in several predictions, including the stellar mass function (SMF) at $z=0$, which lead us to re-tune some of the free parameters.
We re-tuned three parameters associated with AGN feedback {($\kappa$, $\kappa^{}_\mathrm{jet}$, and $\Gamma_\mathrm{thresh}$) and three associated with ejected gas reincorporation ($\tau^{}_\mathrm{reinc}$, $\gamma$, and $M_\mathrm{norm}$) , which we provide and compare to those in \citetalias{lagos2024} in Table \ref{tab:sharktuned}.
For the calibration of these parameters, we use the same approach in \citetalias{lagos2024}, utilising the  built-in optimisation module in \sharkII\ (\texttt{optim}) and calibrating the parameters to the $z=0$ and $z=1$ SMFs of \citet{li2009}.
While this brought most of the prediction roughly in line with those in \citetalias{lagos2024}, we found offsets between PM and L210N1536 in the galaxy-SMBH mass relations and AGN bolometric luminosity functions (the latter being the focus of part II of this work).
To correct these tensions, we manually re-tuned two parameters: $f^{}_\mathrm{SMBH}$ and $e^{}_\mathrm{SB}$, which control the cold gas accretion into the SMBH\footnote{Since cold gas accretion is significantly driven by mergers in \sharkII, differences in halo finding can impact this accretion channel, which dominates both SMBH growth and AGN luminosity.} (see Equations \ref{eq:2} and \ref{eq:3}).
With these choices, we treat the $z=0$ and $z=1$ SMFs as the primary observational constraints that we do our best to quantitively match, and the normalisation of the $z=0$ bulge-SMBH mass relation and AGN bolometric luminosity function as secondary observational constraints that we attempt to qualitatively match without negatively impacting our primary constraints \citep[following the terminology presented by][]{lacey2016}.

\section{Predicted SMBH properties}\label{S3:BH}

\begin{figure}
    \centering
    \includegraphics[width=\linewidth]{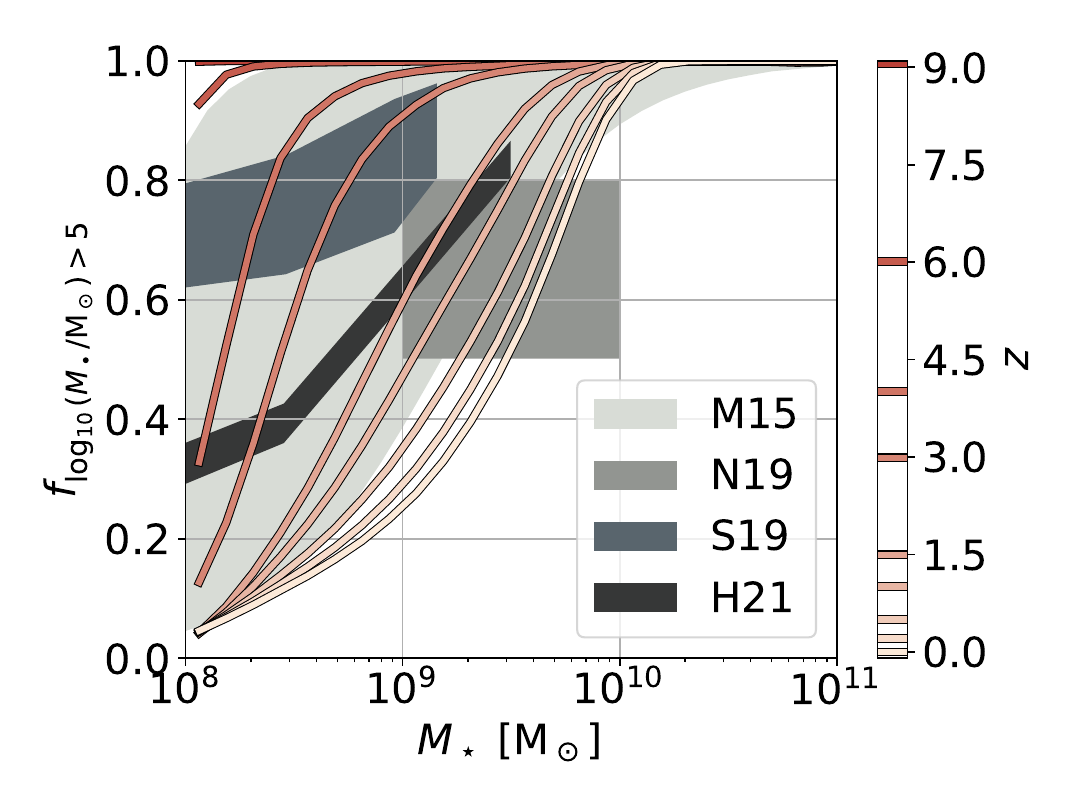}
    \caption{Fraction of galaxies with $\MBH>\msun{5}$ as a function of $M_\star$ at different redshifts. Lines coloured by redshift, with darker lines for higher redshifts.
    We note that almost all galaxies ($>99.999\%$) above $M_\star=\msun{8}$ have a SMBH in the $0\leq z\leq9$ range.
    Inferred SMBH occupation fraction at $z\sim0$ from observations show are shown with shaded areas: from X-ray data \citep{miller2015} in light grey, from dynamical SMBH masses \citep{nguyen2019} in middle grey, from nuclear cluster occupation fractions in the Virgo cluster \citep{sanchez2019} in slate grey, and from nuclear cluster occupation fractions in the Local Volume \citep{hoyer2021} in dark grey.}
    \label{fig:Mstar_fMBH}
\end{figure}

In addition to the stellar mass selection of $M_\star>\msun{8}$, we select and show only galaxies with SMBH masses above $\msun{5}$.
This choice is driven by the significant number of SMBHs with $\MBH\lesssim\msun{5}$ that have grown almost completely by SMBH-SMBH mergers, leading to artificial spikes in the BHMF in masses corresponding to integer multiples of our chosen SMBH mass seed.
This feature is most prominent at $z=0$, lessening towards higher redshifts, with the spikes dissapearing by $z\sim4$.
This is indicative of a fundamental change in the main channel for SMBH growth in \sharkII, from being globally dominated by (gas) accretion at high redshift to a significant contribution from mergers at low redshift.

Figure \ref{fig:BHMF} shows the evolution of the SMBH mass function (BHMF) from $z=9$ to $z=0$.
We note that the apparent peak at $\MBH\sim\msun{6}$ at early times ($z\gtrsim4$) is incompleteness introduced by our stellar mass cut of $M_\star>\msun{8}$, e.g., \sharkII\ predicts that the median $M_\star\sim\msun{8}$ galaxy at $z=9$ has a SMBH of $\sim\msun{6}$.
SMBHs growing to be $\sim1\%$ of their galaxies' total stellar mass is indicative of a highly efficient SMBH growth at early times, which is the reason we find that the choice of SMBH seed mass has only a minor impact in \sharkII, even at $z=9$.
A break in the slope of the BHMF appears by $z\sim4$ at $\sim\msun{8}$ and remains in place down to $z=0$.
The slope of the BHMF above this transition mass shows no strong evolution in the $0<z<4$ range, while slope below does significantly flatten in the same redshift span.
A second break appears in the BHMF by $z\sim1.5$ at $\sim\msun{6}$ and remains in place by $z=0$, below which the slope is steeper.
We also show in Figure \ref{fig:BHMF} the BHMFs for central and satellite galaxies.
While both centrals and satellites show similar breaks at $\sim\msun{8}$, the $\sim\msun{6}$ break is stronger in centrals, indicating that environment can play a role in SMBH growth.

A detailed reconstruction of all measurement uncertainties from observations is outside of the scope of this work, which include a number of assumptions in the conversion from AGN measurements to SMBH masses to BHMF \citep[e.g., see figure 3 of][]{gallo2019}, but for illustration purposes we include also the measured BHMF from \sharkII\ if we assume a universal 0.5 dex error in the measurement of $\MBH$\footnote{Our choice of error is not meant to be an accurate representation of the measurement errors in observations, but as a rough and simple approximation to showcase the possible impact of these errors in the measurement of the BHMF.}.
\sharkII\ roughly matches observational local measurements of the BHMF in the $\MBH\sim10^8$-$\msun{10}$ range, below/above which \sharkII\ under/over-predicts the abundance of SMBHs.
While both ends have larger observational uncertainties due to (low-number statistics) at the low-mass (high-mass) end, we note that an over-prediction at the high-mass end would be expected, given that we over-predict the high-mass end of the SMF (see figure 1 \citetalias{lagos2024}) and our use of the $z=0$ $\Mbulge$-$\MBH$ as secondary calibration.

The match improves in the $0.2\lesssim z\lesssim6.0$ range for SMBH masses above $\sim\msun{8}$, with the observed BHMF from \citet{shankar2009,wu2022b,burke2025} all falling in between the true and error-perturbed BHMFs predicted by \sharkII.
Below $\MBH\sim\msun{8}$, \sharkII\ is not far from the observed BHMF up to $z\sim1.5$, above which it significantly over-predicts the abundance of low-mass SMBH, though we note that observationally measuring the BHMF in this mass and redshift range poses significant challenges.
While less constraining, we also find that the \sharkII\ BHMF remains fully consistent with the observational measurements limited to specific subsets of galaxies/AGN from \citet{shen2012,ananna2022,he2024,matthee2024} across cosmic time.
Overall, we find that \sharkII\ produces a BHMF that is in reasonable agreement with those found in the observational literature, with a prediction that a significant number of low-mass SMBHs might be found by future deeper surveys at high redshift ($z\gtrsim3$).

\begin{figure*}
    \centering
    \includegraphics[width=\linewidth]{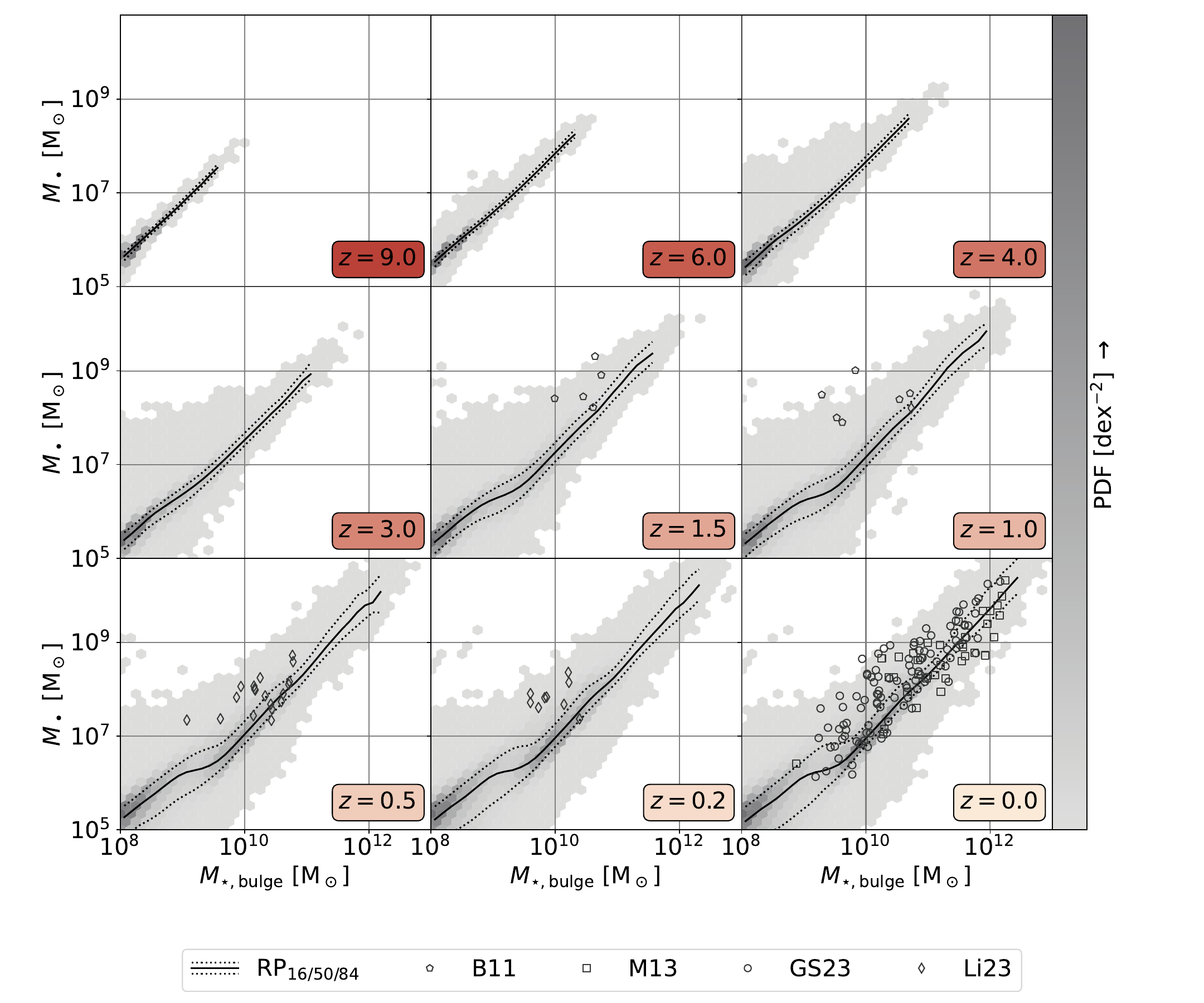}
    \caption{Comparison of the predicted $\Mbulge$-$\MBH$ distribution in \sharkII\ from $z=9$ to $z=0$ to observational results from the literature.
    Each panel corresponds to the $\Mbulge$-$\MBH$ distribution at the redshift indicated at the bottom right corner.
    The coloured histograms in the background show the overall distribution of \sharkII\ galaxies at each redshift, linearly coloured and with the scale adjusted for maximum contrast in each panel (i.e., the exact value of a given shade changes from panel to panel), with higher density regions in darker shades.
    The solid and dotted lines showing the running medians and 16\nth/84\nth\ running percentiles, respectively.
    Literature measurements from observations are shown in panel with the closest redshift.
    Observational data shown: \citet{bennert2011} with pentagons, \citet{mcconnell2013} with squares, \citet{graham2023a} with circles (including the \citealt{chabrier2003} IMF correction from \citealt{graham2024}), and \citet{li2023b} with diamonds.}
    \label{fig:Mbulge_MBH}
\end{figure*}

While the vast majority of galaxies with $M_\star\geq\msun{8}$ have been seeded with a SMBH in \sharkII\ ($>99.999\%$), irrespective of redshift, not all galaxies provide the necessary conditions for their SMBHs to grow.
Figure \ref{fig:Mstar_fMBH} shows the fraction of galaxies with $\MBH>\msun{5}$ as a function of stellar mass and redshift, which shows clear trends with both.
At $z=9$, most galaxies ($\gtrsim90\%$) have a SMBH with a mass of at least $\msun{5}$, with the fractions declining towards $z=0$ at $M_\star$-dependent rates, with low-mass galaxies seeing the largest change with redshift (from $\sim90\%$ at $z=9$ to $\lesssim10\%$ by $z=3$).
This is a clear indication of a reduction in SMBH masses at fixed stellar masses towards $z=0$  (i.e., a downward shift in the $M_\star$-$\MBH$ relation, as shown in \ref{S3.1:Mbulge_MBH} and \ref{S3.2:Mstar_MBH}).

Compared with measurement of the SMBH occupation fraction from $z\sim0$ observations, the fraction of galaxies with $\MBH>\msun{5}$ predicted by \sharkII\ lies on the lower end of measurements, though we caution against over-interpretation of this result as the observational measurements are not consistent across different methods and that they do not include the same explicit limit on $\MBH$ as we do.
Most illustrative is the comparison with the results from \citet{burke2025}, which we omit from Figure \ref{fig:Mstar_fMBH} for clarity.
Our BHMF lies on the high end of their measured BHMF (see Figure \ref{fig:BHMF}), which would imply an occupation fraction of $\sim1$, which is what \citet{burke2025} find for galaxies with $M_\star\gtrsim\msun{8}$ but appears inconsistent with our results in Figure \ref{fig:Mstar_fMBH}.
The extend BHMF they present suggests that they probe into much lower BH masses \citep[$\MBH\sim\msun{2}$, see figure 12 of][]{burke2025} than we do, and if we include all seeded SMBHs in \sharkII\ we match their $\sim1$ occupation fraction, highlighting the challenge of comparing to observational measurements with no clear SMBH mass selection cuts.

\subsection{The $\Mbulge$-$\MBH$ relation}\label{S3.1:Mbulge_MBH}

\begin{figure*}
    \centering
    \includegraphics[width=\linewidth]{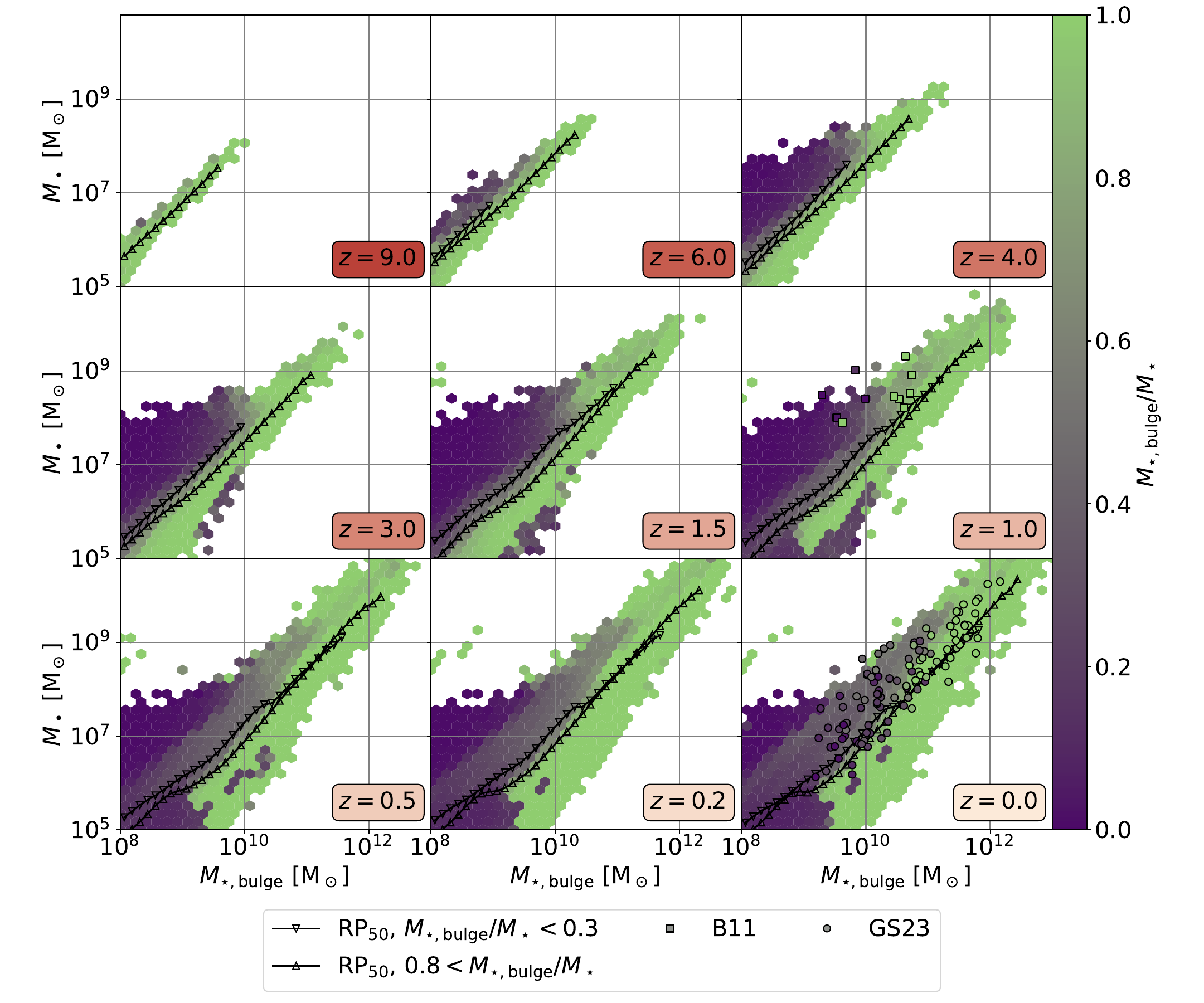}
    \caption{Similar to Figure \ref{fig:Mbulge_MBH}, coloured by the median bulge-to-total mass ratio ($\Mbulge/M_\star$) of the galaxies in each cell, and with two lines showing only the running medians for early-type ($\Mbulge/M_\star>0.8$, shown with triangle markers pointing upwards) and late-type galaxies ($\Mbulge/M_\star<0.3$, shown with triangle markers pointing downwards).
    Observational data shown: \citet{bennert2011} with squares and \citet{graham2023a} with circles, both coloured according to $\Mbulge/M_\star$ of each galaxy (with the same colour scale used for our \sharkII\ data).}
    \label{fig:Mbulge_MBH_BT}
\end{figure*}

Figure \ref{fig:Mbulge_MBH} shows the predicted redshift evolution in the $\Mbulge$-$\MBH$ plane.
\sharkII\ produces a tight correlation between $\Mbulge$ and $\MBH$ across cosmic time, though galaxies below $\Mbulge\sim\msun{10}$ can be strong outliers with overly/under-massive SMBHs by up to $\sim2$ dex already by $z=3$.
The scatter around the median relation grows towards $z=0$ with a strong bulge stellar mass dependence (see Appendix \ref{A1:MBH_scatter} for further details).
The break in the BHMF at $\MBH\sim\msun{6}$ is apparent in this relation as a change in the slope in the median relation, which roughly corresponds to the bulge mass where the scatter in the relations significantly decreases (see Appendix \ref{A1:MBH_scatter}).
While the excellent agreement between \sharkII\ and $z\sim0$ observations is partially by construction as we calibrate the free parameters to roughly reproduce the observed normalisation (as mentioned in \ref{S2.2:PM}), we made no effort to match the slope of the relation at $z=0$, which is a true prediction from the model.
The agreement with observations remains excellent up to $z=0.5$, and even at higher redshifts \sharkII\ is able to populate most of the space occupied by observations.

\begin{figure*}
    \centering
    \includegraphics[width=\linewidth]{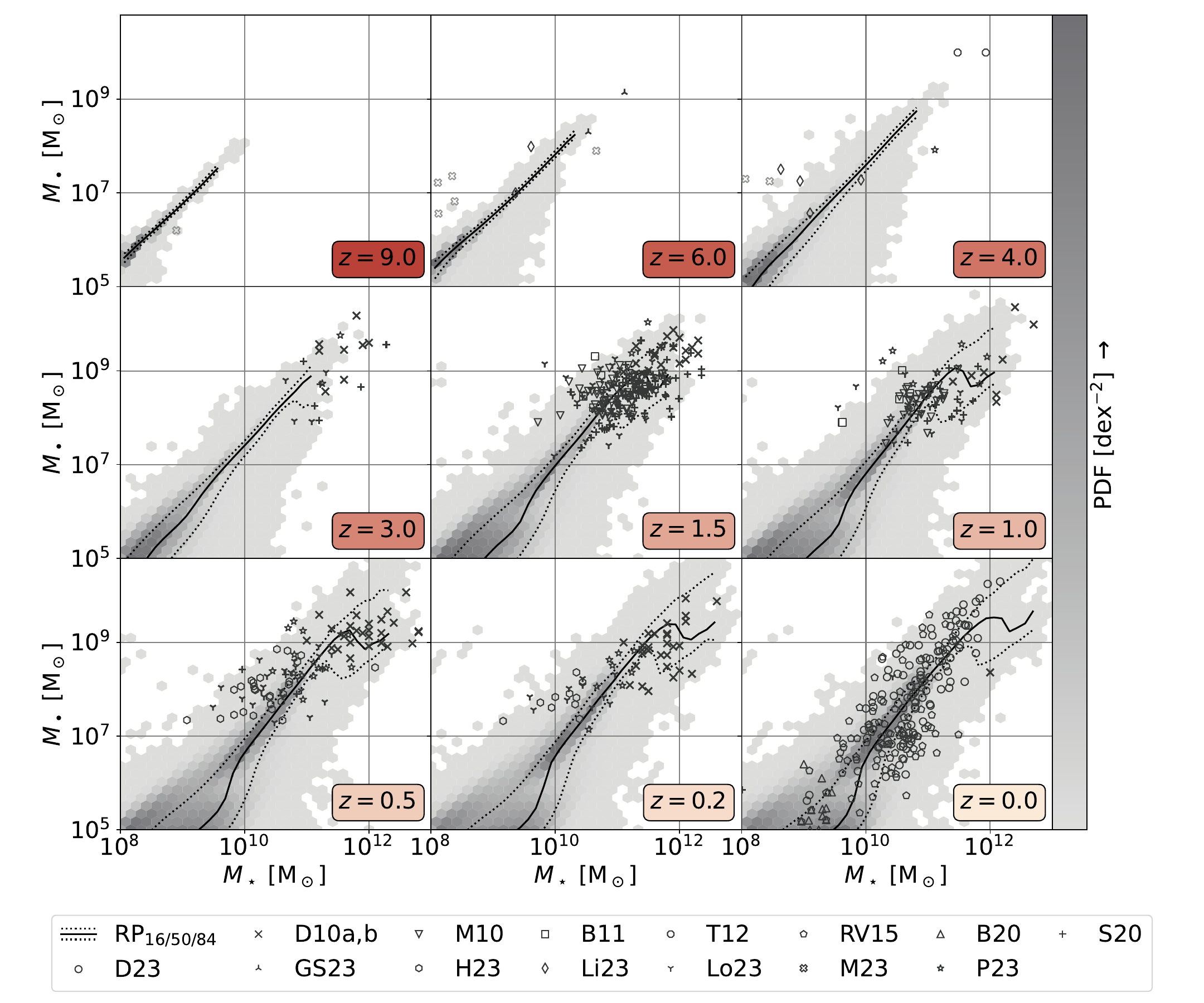}
    \caption{Similar to Figure \ref{fig:Mbulge_MBH}, showing the comparison of the predicted $M_\star$-$\MBH$ distribution in \sharkII\ from $z=9$ to $z=0$ to observational results from the literature.
    Observational data shown: \citet{decarli2010a,decarli2010b} with crosses, \citet{merloni2010} with downward triangles, \citet{bennert2011} with squares, \citet{targett2012} octagons, \citet{reines2015} with pentagons, \citet{baldassare2020} with upward triangles, \citet{suh2020} with thin crosses, \citet{ding2023} with circles, \citet{graham2023a} with three-pointed stars, \citet{harikane2023} with hexagons, \citet{li2023b} with diamonds, \citet{lopez2023} with inverted three-pointed stars, \citet{maiolino2024} with thick crosses, and \citet{poitevineau2023} with stars.}
    \label{fig:Mstar_MBH}
\end{figure*}

Recent works have shown evidence that the $z=0$ relation between SMBH and bulge stellar masses depends on galaxy morphology, with early- and late-type galaxies following different relations \citet[e.g.,][]{savorgnan2016,dullo2020,graham2023a}.
Using the distribution of bulge-to-total ratios ($\Mbulge/M_\star$) of the early- and late-type galaxies in \citet{graham2023a} as a rough guide to classify our galaxies as early- or late-type, Figure \ref{fig:Mbulge_MBH_BT} shows that separating \sharkII\ galaxies by morphology leads to different relations, with $z=0$ late-type galaxies having more massive SMBHs than early-types at fixed $\Mbulge$ with a roughly constant offset (we expand this comparison to the fits provided by \citealt{graham2023a} in Appendix \ref{A2:GS23_comp}).
The few $z\sim1$ measurements from \citet{bennert2011} also show a qualitative agreement with our predictions. 
We find that separation between the median relations of early- and late-type galaxies is well-established by $z\sim4$.
Outliers from the overall relations are for the most part dominated by one of the two morphological types, with galaxies with $\Mbulge\lesssim\msun{10}$ and overly-massive SMBHs mostly being late-type and the rest of outliers being mostly early-type galaxies.
Regardless of redshift, \sharkII\ predicts that the most SMBHs are always found in early-type galaxies.

We opt not to show the difference between centrals and satellites in this relation as they only show minor differences.
Both populations are nearly undistinguishable from $z=9$ down to $z=3$, where small differences start to appear at low bulge masses ($\lesssim\msun{9}$).
The differences become more pronounced by $z=0.5$, where the satellites exhibit a narrower distribution of SMBH masses for $\Mbulge\lesssim\msun{10}$ and a stronger break in the relation at $\Mbulge\lesssim\msun{9}$.
In contrast, satellites with massive bulges ($\Mbulge\gtrsim\msun{11}$) remain consistent with central galaxies by $z=0$.

\subsection{The $M_\star$-$\MBH$ relation}\label{S3.2:Mstar_MBH}

\begin{figure*}
    \centering
    \includegraphics[width=\linewidth]{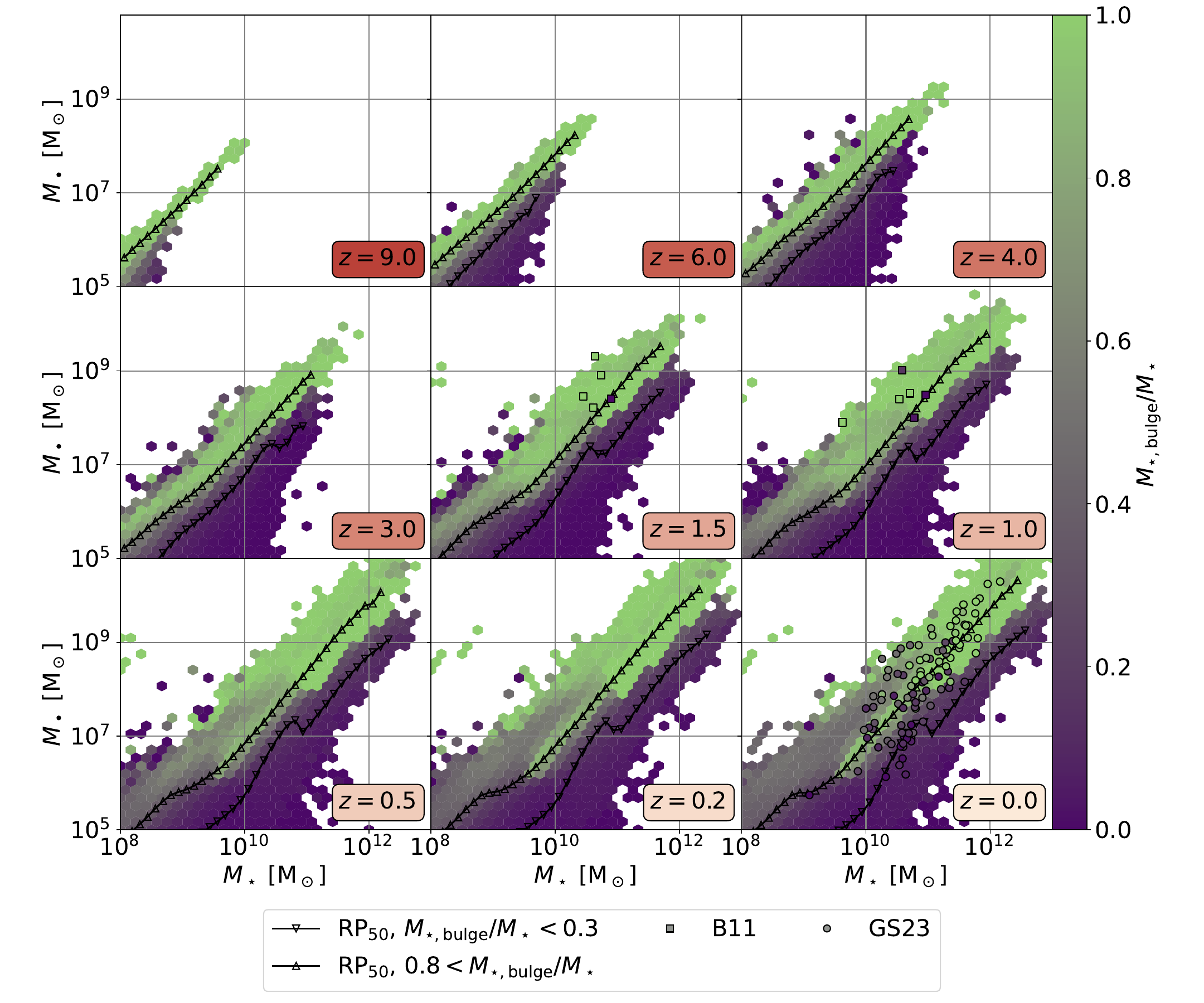}
    \caption{Similar to Figure \ref{fig:Mstar_MBH}, coloured by the median $\Mbulge/M_\star$ of the galaxies in each cell, as in Figure \ref{fig:Mbulge_MBH_BT}.
    Observational data shown: \citet{bennert2011} with squares and \citet{graham2023a} with circles (including the \citealt{chabrier2003} IMF correction from \citealt{graham2024}).}
    \label{fig:Mstar_MBH_BT}
\end{figure*}

Examining the relation between total stellar mass and SMBH mass, we find the same trend of, at fixed stellar mass, decreasing SMBH masses towards $z=0$, as shown in Figure \ref{fig:Mstar_MBH}.
The main difference between the two relations is in the scatter, with the scatter in the $M_\star$-$\MBH$ relation being significantly larger than the $\Mbulge$-$\MBH$ equivalent, another trait well-established in the literature \citep[e.g.,][]{kormendy2013,graham2023a}.
The dispersion around the $M_\star$-$\MBH$ relation shows a more complex dependence with stellar mass and time evolution, with two local maxima that move to higher stellar masses from high-$z$ towards $z=0$ (see Appendix \ref{A1:MBH_scatter} for more details).
Overall, we find that \sharkII\ is mostly able to produce galaxies that correspond to observational measurements of $M_\star$ and $\MBH$.

\begin{figure*}
    \centering
    \includegraphics[width=\linewidth]{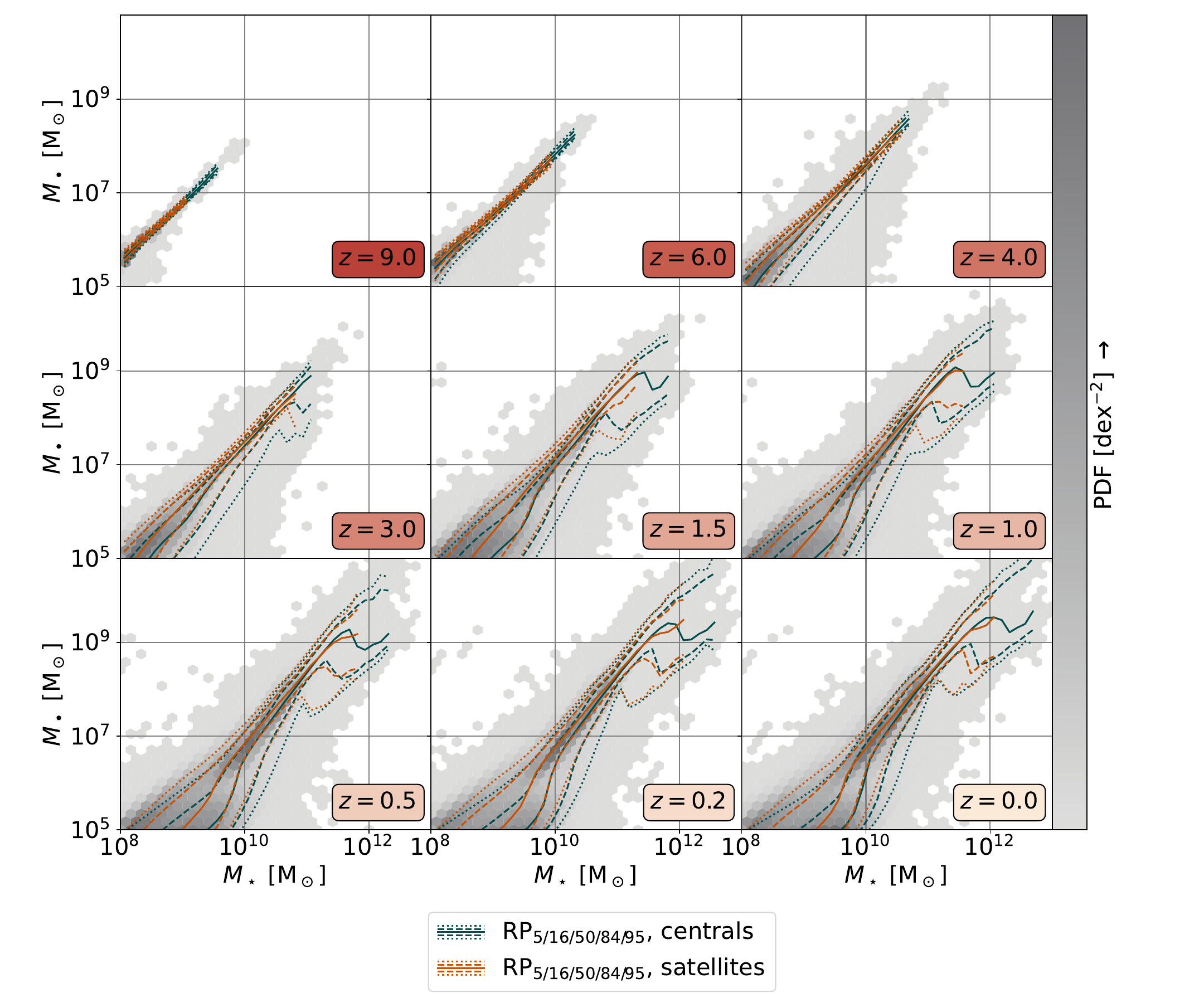}
    \caption{Similar to Figure \ref{fig:Mstar_MBH}, showing on the $M_\star$-$\MBH$ relation for central and satellite galaxies.
    We have omitted the observational measurements shown in Figure \ref{fig:Mstar_MBH} for clarity.}
    \label{fig:Mstar_MBH_censat}
\end{figure*}

While we note that observations suggest a flatter $M_\star$-$\MBH$ relation at $z\gtrsim4$ than what \sharkII\ predicts, this can easily happen if a mild bias is present in the observations which prefers the detection of more massive black holes.
Overly massive SMBHs are overrepresented due to being brighter at fixed Eddington ratio \citep[known as Lauer bias;][]{lauer2007}, which may particularly impact low-mass galaxies which tend to have very low accretion ratios \citep[e.g.,][]{urquhart2022}.
Such a stellar mass-dependent bias would naturally lead to a flatter measured observations.
In addition, outside of comparatively few local galaxies, the measurement of SMBH masses depends on them being active enough for detection and measurement, which can further bias the measurements if the SMBH of AGN-selected galaxies are not representative of the overall galaxy population.
The lack of agreement on evolutionary trends \citep[e.g.,][]{merloni2010,shen2015,ueda2018,suh2020,ding2020} and possible biases for the SMBH mass measurements at higher redshifts \citep[e.g.,][]{ananna2024,lupi2024} also complicate the interpretation of our results of a strong evolution in the SMBH scaling relations.

\begin{figure*}
    \centering
    \includegraphics[width=\linewidth]{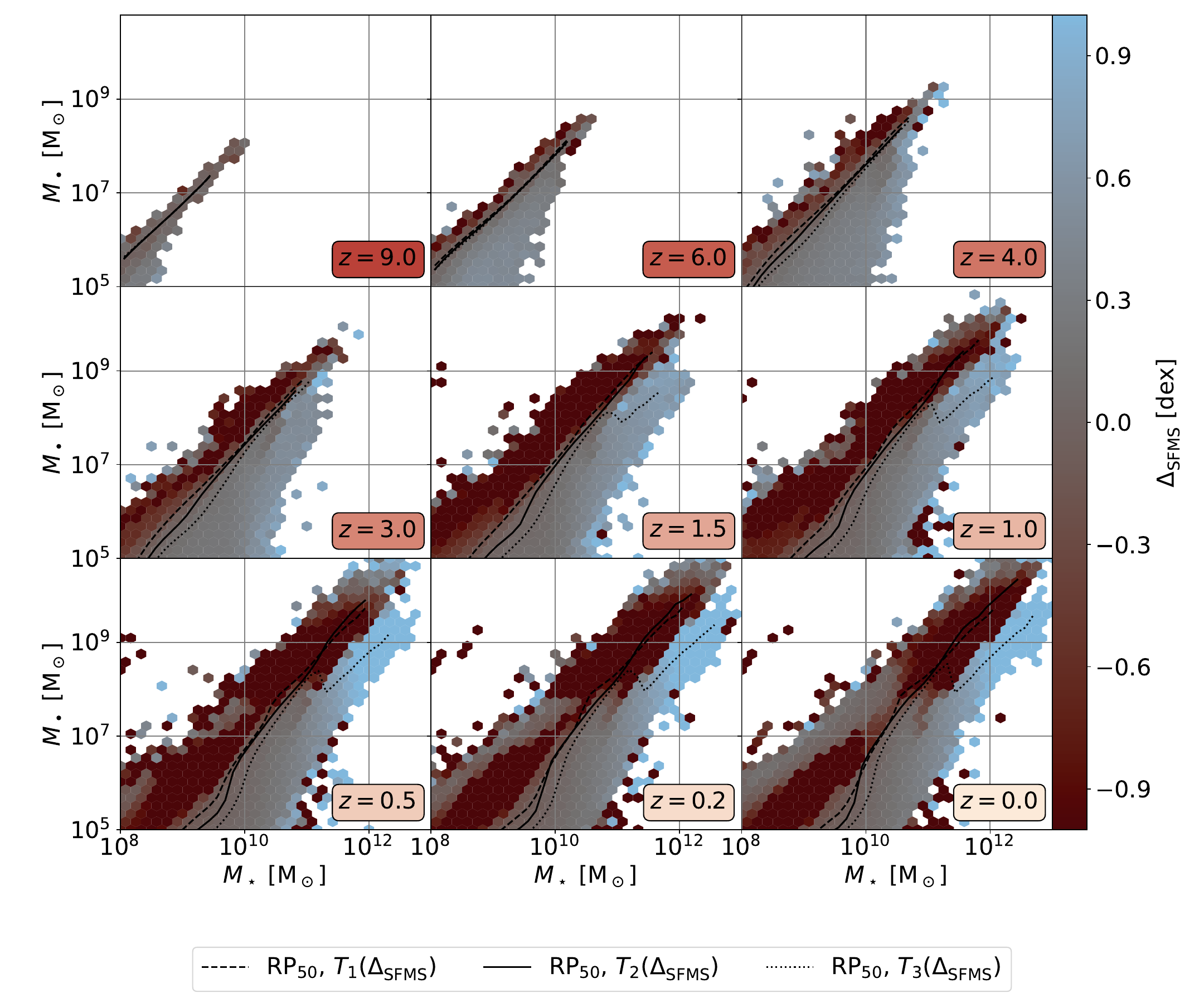}
    \caption{Similar to Figure \ref{fig:Mbulge_MBH_BT}, coloured by the offset from the star-forming main sequence in \sharkII\ ($\DeltaSFMS$).
    The three black lines indicate the running median for galaxies divided into $\DeltaSFMS$ terciles, with the lowest (highest) tercile, $T_1$ ($T_3$) being the one with the lowest (highest) $\DeltaSFMS$ values (i.e., $T_1$ roughly corresponds to galaxies below the SFMS and $T_3$ with galaxies above the SFMS).}
    \label{fig:Mstar_MBH_SFMS}
\end{figure*}

In line with recent observational results \citep[e.g.,][]{savorgnan2016,dullo2020,graham2023a}, we find that bulge-dominated (early-type) and disc-dominated (late-type) galaxies follow distinct relations in the $M_\star$-$\MBH$ plane at $z\sim0$ in \sharkII, as shown in Figure \ref{fig:Mstar_MBH_BT}.
This separation is already apparent at $z\sim9$, with both populations becoming clearly distinct by $z\sim6$, with disc-dominated galaxies exhibiting systematically lower median SMBH masses than bulge-dominated galaxies at fixed stellar mass (and with a larger separation than in the $\Mbulge$-$\MBH$ plane).
This morphological split is also apparent in the most extreme outliers from the overall relation, with disc-dominated and bulge-dominated galaxies occupying different areas of the plane.
We find that the predictions from \sharkII\ are in good qualitative agreement with the overall location of bulge- and disc-dominated galaxies from the observational literature (we also compare to the fits provided by \citealt{graham2023a} in Appendix \ref{A2:GS23_comp}).

\begin{figure*}
    \centering
    \includegraphics[width=\linewidth]{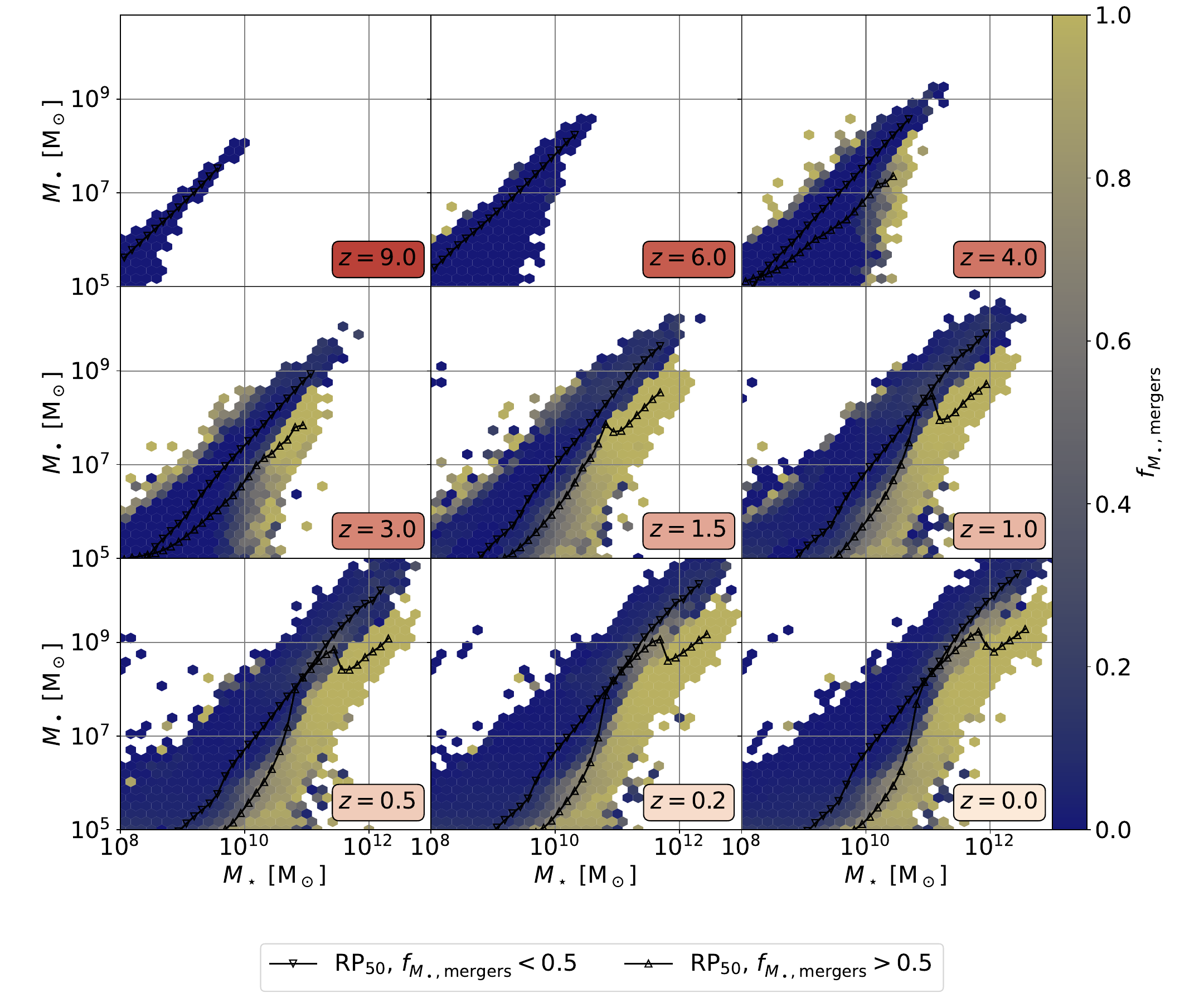}
    \caption{Similar to Figure \ref{fig:Mstar_MBH_SFMS}, coloured by the fraction of the SMBH mass built through SMBH-SMBH mergers in \sharkII\ ($\fmerge$), and with two lines showing only the running medians for SMBHs built mostly by mergers ($\fmerge>0.5$, shown with triangle markers pointing upwards) and those built mostly by gas accretion ($\fmerge<0.5$, shown with triangle markers pointing downwards).}
    \label{fig:Mstar_MBH_fmerge}
\end{figure*}

To explore the effect of environment in the $M_\star$-$\MBH$ relation, in Figure \ref{fig:Mstar_MBH_censat} we split the \sharkII\ galaxy population between central and satellite galaxies.
We find that differences are already present at $z\sim4$, with satellite galaxies below $\sim\msun{10}$ exhibiting more massive SMBHs than centrals of the same stellar mass, with the difference reaching $\sim0.5$ dex by $z=0$.
Galaxies above $\sim\msun{10}$ exhibit little-to-no difference when divided between centrals and satelllites, indicating that \sharkII\ predicts the SMBH growth above this stellar mass to be dominated by non-environmental processes, while environment plays a significant role for satellite galaxies below this stellar mass.

We now explore the connection between these relations, with a focus on the $M_\star$-$\MBH$ relation, and other $M_\star$-based scaling relations, in particular the star-forming main sequence (SFMS) and stellar mass-size relation (SMSR).
For this purpose, we fit each relation as described in Appendix \ref{A3:mass_relations}, measure the offset of each galaxy from the fits at their stellar mass, and then explore trends in the $M_\star$-$\MBH$ plane.

Starting with the SFMS, we show the median offset ($\DeltaSFMS$) in the $M_\star$-$\MBH$ plane in Figure \ref{fig:Mstar_MBH_SFMS}, where positive (negative) values indicate that the median galaxy lies above (below) the SFMS, together with running medians for galaxies divided by $\DeltaSFMS$ terciles.
We find a clear trend at redshifts $0.5\lesssim z\lesssim6$, with galaxies that lie below the SFMS having oversized SMBHs relative to galaxies on/above the SFMS of similar stellar mass.
At $z=0.5$ a more complex separation starts to appear, with the impact of the offset from the SFMS being less significant around $\sim\msun{10}$, the most extreme outliers below this mass being mostly SF galaxies and passive galaxies dominating just above the main relation.
Only above $M_\star\sim\msun{10}$ does the anti-correlation between offsets from the SFMS and $M_\star$-$\MBH$ relation remains at $z=0$.
Some differences are also apparent in the median $\MBH$ when galaxies are divided in $\DeltaSFMS$ terciles, but these are comparatively small effects compared to morphology.
Overall, it is clear that \sharkII\ predicts a strong connection between star formation and SMBH mass, in particular with large SMBHs being connected to the quenching of star formation, in qualitative agreement with recent observational results \citep[e.g.,][]{piotrowska2022,bluck2023,graham2023c}.

While we do not show either in this work, we have also explored the connection between the $M_\star$-$\MBH$ and offset from three more scaling relations: stellar mass-size relation (SMSR) defined using the half-mass radii ($r^{}_{50,\star}$), and (host) halo-stellar mass relation for both central and satellite galaxies separately.
Trends with the SMSR offset appear by $z\sim4$, with under/over-sized galaxies having over/under-sized SMBHs for their stellar mass down to $z\sim1$, where the median galaxy along the SMSR become the one hosting the smallest SMBHs.
These trends indicate that, at fixed stellar mass, SMBHs grow to larger masses in the more compact galaxies in the redshift range $1\lesssim z \lesssim4$, while at lower redshifts there is a change to SMBH growth being boosted by processes that move galaxies away from the SMSR in either direction (i.e., either more compact or extended than normal).
In comparison, the offset from both halo mass relations show no clear correlation with the scatter in both $\Mbulge$-$\MBH$ and $M_\star$-$\MBH$ relations, indicating that halo mass plays no significant role in shaping the scatter in these relations.

Inspired by the discussion presented in \citet{graham2023a,graham2023b} on the role of wet and dry mergers driving the morphological dependence of the $\Mbulge$-$\MBH$ and $M_\star$-$\MBH$ relations, who show that dry mergers of galaxies in the late-type relation can explain the formation of the observed separate relation for early-type galaxies, we now explore whether gas accretion or SMBH mergers drive the growth of SMBH in \sharkII.
Figure \ref{fig:Mstar_MBH_fmerge} shows how the $M_\star$-$\MBH$ relation depends on the fraction of $\MBH$ that comes from SMBH-SMBH mergers, $\fmerge=M_{\bullet,\mathrm{mergers}}/\MBH$, as a function of cosmic time.
The build-up of SMBHs is dominated by gas accretion at early times ($z\gtrsim6$), after which SMBH-SMBH mergers have a more significant role.
Dividing galaxies between those with SMBH built mostly through gas accretion ($\fmerge<0.5$) and those built mostly through mergers ($\fmerge>0.5$), we find that significant differences are already apparent by $z\sim4$, becoming more pronounced towards $z=0$.
For most galaxies, those low $\fmerge$ exhibit systematically larger median $\MBH$ than those with high $\fmerge$, which is also apparent in the most extreme outliers of the $M_\star$-$\MBH$ relation, with the former (latter) being responsible for the most over-massive (under-massive) SMBHs at fixed stellar mass.
Galaxies with $M_\star\sim\msun{11.5}$ from $z\sim1$ downwards are partial exception to this trend, as the median $\MBH$ does not depend on $\fmerge$, with the stellar mass range with small difference in the medians widening towards $z=0$ (where it spans the $\sim1$ dex in stellar mass).

We note that we see different trends in the $\Mbulge$-$\MBH$ relation with $\fmerge$ (not shown for brevity).
We also find a clear separation of $\sim0.5$ dex at $z\sim4$ between SMBHs built mostly through gas accretion or mergers, with the latter having systematically higher $\MBH$ at fixed $\Mbulge$ (similar to the inversion of the trend with morphology between $\Mbulge$-$\MBH$ and $M_\star$-$\MBH$ relations), though this offset disappears by $z\sim1.5$.
By $z\sim0.5$ a new offset appears, reversing the previous trend (also by $\sim0.5$ dex), but only for galaxies with $\Mbulge\lesssim\msun{9.5}$, while the medians of both populations remain similar above this stellar mass.
The extreme outliers do show a trend, with galaxies with high $\fmerge$ making most of the outliers and those with low $\fmerge$ mostly dominating the close to the median relation.
While we are not able to reproduce the more refined morphological classification used by \citet{graham2023a,graham2023b} to directly compare the evolutionary picture they presented, we find that \sharkII\ does produce a similar scenario where dry and wet mergers (leading to high and low $\fmerge$, respectively) shape distinct scaling relations in the $M_\star$-$\MBH$ plane, and to a lesser degree in the $\Mbulge$-$\MBH$ plane.

\section{Discussion}\label{S4:disc}

We start this section with our interpretation of the evolution of the SMBH scaling relations as predicted by \sharkII.
At early times ($z\gtrsim6$), SMBH growth is dominated by gas accretion (Figure \ref{fig:Mstar_MBH_fmerge}), which is consumed fast enough to lead to the formation of a clear peak at $\MBH\sim\msun{6}$ in the BHMF already by $z=9$ (Figure \ref{fig:BHMF}), growth that is tightly connected to the growth of the stellar mass in the galaxy leading to small scatter in both $\Mbulge$-$\MBH$ and $M_\star$-$\MBH$ relations.
Since cold gas contributes most of the mass growth due to gas accretion in \sharkII, which is driven by either disc instabilities or galaxy mergers, it is not surprising that this rapid growth phase is strongly connected with the formation and growth of a significant spherical stellar component from the accompanying starburst (Figure \ref{fig:Mstar_MBH_BT}).
By $z\sim4$ changes in the evolution of galaxies and their SMBHs start to become visible, with a trend of galaxies hosting smaller SMBHs at fixed bulge/total stellar mass towards $z=0$, coupled with an increase in the scatter around both relations (Figures \ref{fig:Mbulge_MBH}, \ref{fig:Mstar_MBH}, and \ref{fig:Mstar_DMBH}).

A divide between multiple populations also appears at $z\sim4$, with satellite and quenched galaxies starting to show higher $\MBH$ compared to central and SF galaxies, respectively (Figures \ref{fig:Mstar_MBH_censat}, \ref{fig:Mstar_MBH_SFMS}, and \ref{fig:Mstar_MBH_fmerge}).
All of these trends become more pronounced towards $z=0$, with some also developing more complex relation with other galaxy properties (like morphology or star formation).
In particular, the large increase in scatter towards $z=0$ seen in the $M_\star$-$\MBH$ ($\Mbulge$-$\MBH$) relation for $10^9\lesssim M_\star/\mathrm{M}_\odot \lesssim10^{10}$ and $10^{11}\lesssim M_\star/\mathrm{M}_\odot \lesssim10^{12}$ ($\Mbulge\lesssim\msun{9.5}$) indicates an increasing variety of paths for galaxies and their SMBHs.
These differences appear strongly connected to how the galaxies have built their SMBHs, with $\MBH$ of those built mostly through gas accretion being systematically different from those built from SMBH-SMBH mergers.
Combined, this presents a picture where the details of the mergers that galaxies experience, i.e., whether the merger funnels gas into the centre of the galaxy to fuel the growth of the SMBH, dictates the evolution of galaxies through the $\Mbulge$-$\MBH$ and $M_\star$-$\MBH$ planes and shape the overall relations.

\begin{figure}
    \centering
    \includegraphics[width=\linewidth]{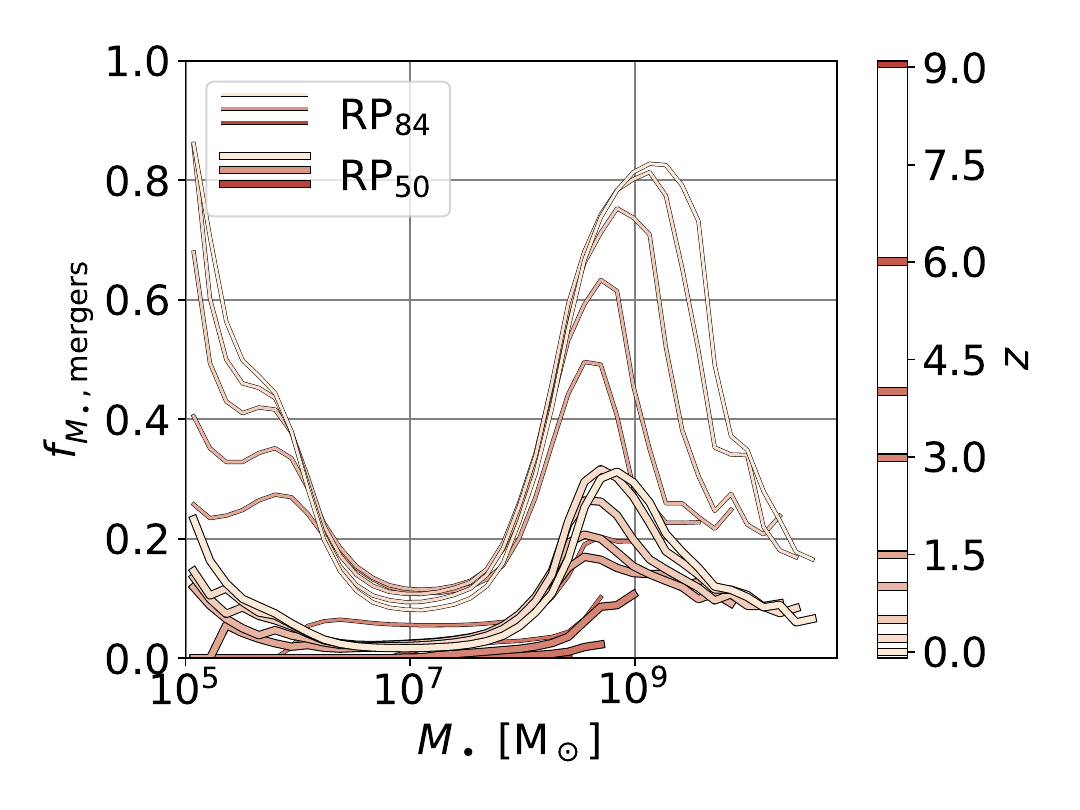}
    \caption{Fraction of the SMBH mass built through SMBH-SMBH mergers, $f^{}_{\MBH,\mathrm{mergers}}$, as a function of $\MBH$ and redshift.
    Thick lines indicate the running medians, and the thin lines the running 84\nth\ percentiles (running 16\nth\ percentiles not shown for clarity).
    Lines are coloured by redshift as in Figure \ref{fig:Mstar_fMBH}.}
    \label{fig:MBH_fmerge}
\end{figure}

Something that is not necessarily apparent from the results we have shown up to this point is that $\fmerge$ shows a remarkable dependency with $\MBH$, and by extension with $\Mbulge$ (also with $M_\star$ to a lesser degree).
Figure \ref{fig:MBH_fmerge} shows that, while there is an overall trend to an increased contribution of SMBH-SMBH mergers from $z=9$ to $z=0$, SMBHs with $\MBH\sim\msun{7}$ show little change on how their mass has been built over $\sim13$ Gyr.
We find a similar trend for $\fmerge$ as function of $\Mbulge$, with the minimum in $\fmerge$ at $\Mbulge\sim\msun{10}$, which is not surprising given the tight relation between it and $\MBH$.
We also find a roughly similar trend in $\fmerge$ with $M_\star$, with a narrower minimum at $\sim\msun{10.5}$.
This feature is even more intriguing because it does not seem to correlate to any other feature in either relation, outside of the change in the scatter in the $\Mbulge$-$\MBH$ relation (as shown in Appendix \ref{A1:MBH_scatter}).
We leave the exploration of how the different prescriptions in \sharkII\ combine to predict that the growth of $\MBH\sim\msun{7}$ SMBHs in galaxies with $\Mbulge\sim\msun{10}$ ($M_\star\sim\msun{10.5}$) mostly grow their SMBHs through gas accretion even by $z=0$, along with a more detailed exploration of how galaxies move across these relations and the drivers of that evolution, to future work.

\subsection{Disentangling the drivers of the scatter in the $\Mbulge$-$\MBH$ and $M_\star$-$\MBH$ relations}\label{S4.1:RFR}

Inspired by the recent use of machine learning techniques to explore the correlation between one galaxy property to a number of other properties \citep[e.g.,][]{piotrowska2022,bluck2023,goubert2024}, we apply a similar method to \sharkII\ to disentangle the effect of these different galaxy properties in the $M_\star$-$\MBH$ and $\Mbulge$-$\MBH$ relations.
To this end, we use the random forest regression \citep[e.g.,][]{amit1997,ho1998,breiman2001} implementation in \textsc{scikit-learn}.
We measure for each galaxy the offset from each relation at their given bulge/total stellar mass and redshift, i.e., how over- or under-massive their SMBH is relative to the median relation, which we use as our target values for the random forest.

\begin{figure*}
    \centering
    \includegraphics[width=\linewidth]{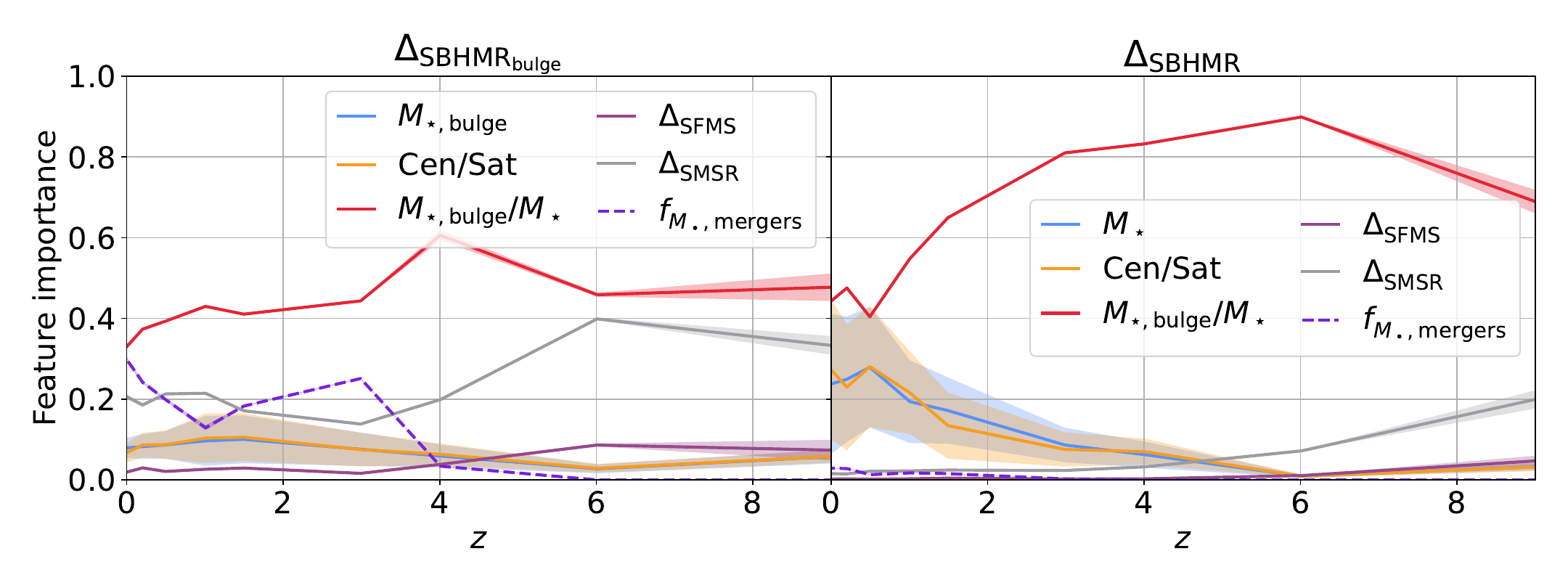}
    \caption{Relative importance of different galaxy properties in explaining the offset from the $\Mbulge$-$\MBH$ (left column) and $M_\star$-$\MBH$ (right column) relations as a function of redshift, measured as the feature importance from applying Random Forest Regressions to the data at each snapshot.
    The lines (shaded areas) show the relative importance (standard deviation of the importance) of each galaxy property: $\Mbulge$ or $M_\star$ (left and right columns respectively, blue), central/satellite classification (orange), $\Mbulge/M_\star$ (red), offset from the star-forming main sequence ($\DeltaSFMS$, magenta), offset from the stellar mass-size relation ($\DeltaMSR$, grey) , and the fraction of the SMBH mass built through SMBH-SMBH mergers ($\fmerge$, dashed purple).}
    \label{fig:FeatureImportance_MB}
\end{figure*}

For input for the regressor we choose six of the galaxy properties we have explored so far in this work: $\Mbulge$ and $M_\star$ (each for their respective scaling relations), $\Mbulge/M_\star$, $\DeltaSFMS$, $\DeltaMSR$, $\fmerge$, and central/satellite classification.
Our choice of using $\DeltaSFMS$ and $\DeltaMSR$ instead of SFR (or sSFR) and galaxy size, respectively, is to account for the strong relations between them and $M_\star$.
In comparison, both $\Mbulge/M_\star$ and $\fmerge$ show a significantly larger scatter as a function of either $\Mbulge$ or $M_\star$.
To avoid biasing the random forest towards the more numerous central galaxies, we reduce the our galaxy catalogue in each snapshot to a (stellar) mass-matched sample with equal numbers of centrals and satellites (including orphans\footnote{As in the \sharkII\ outputs, we use a value of 0 for central galaxies and 1 for satellites.
We set the value for orphan galaxies, which are identified in the outputs with a value of 2, also to 1 as they are treated nearly the same as satellites in terms of physical prescriptions controlling their evolution.}).

Figure \ref{fig:FeatureImportance_MB} shows the resulting relative feature importance from the random forest regressors fit to our \sharkII\ galaxies.
While there are quantitative differences between both relations and with the inclusion or exclusion of $\fmerge$ from the set of features in the relative importance values for the scatter, they show a number of shared trends: $\Mbulge/M_\star$ is the dominant feature across cosmic time, $\DeltaSFMS$ always plays a small role in predicting the offset from both SMBH relations, and that $\DeltaMSR$ is the second most important feature for $z\gtrsim4$.
Below $z\sim4$, both $\DeltaMSR$ and $\fmerge$ have similar secondary importance explaining the scatter in the $\Mbulge$-$\MBH$ relation, while both $M_\star$ and central/satellite classification are the secondary features for the $M_\star$-$\MBH$ relation.

All the properties that account for most of the scatter are intrinsic properties of the galaxies, which, combined with the small importance of the central/satellite classification, clearly show that \sharkII\ predicts that SMBHs masses are not directly influenced by the galaxy environment.
Hence, the overly-massive SMBHs that \sharkII\ predicts for satellite galaxies (Figure \ref{fig:Mstar_MBH_censat}) are an indirect effect of environment-driven changes of the intrinsic properties of the galaxies, with the SMBH co-evolving with these changes.
This indirect impact means that we would expect the difference between centrals and satellites to disappear once these scaling relations are controlled for other intrinsic properties (mainly morphology), in line with previous theoretical predictions in the literature \citep[e.g.,][]{ricarte2019}.
Overall, these results indicate that the physical drivers of galaxy morphology also dictate the evolution of these relations, with other properties playing a secondary role or being correlated to SMBHs through their correlations with morphology. 
In turn, this suggests that the addition of stellar masses and sizes could be used to improved the observational inference of SMBH masses based on the $\Mbulge$-$\MBH$ relation, and that inferences purely based on $M_\star$ would be hard to improve lacking $\Mbulge$ measurements (at which point it becomes redundant).

\subsection{Comparison with predictions from other theoretical models}\label{S4.2:sims}

Up to this point, we have focused on comparing \sharkII\ with observations whenever available, but it is important to explore the differences with other theoretical models.
We focus on the work by \citet{habouzit2021,habouzit2022}, where they explored the evolution of the $M_\star$-$\MBH$ relation in six different hydrodynamical simulations: Illustris \citep{vogelsberger2014}, TNG100 and TNG300 \citep{pillepich2018,springel2018}, Horizon-AGN \citep{dubois2014}, EAGLE \citep{crain2015,schaller2015}, and SIMBA \citep{dave2019}.
All six predict that the $M_\star$-$\MBH$ relation evolves with cosmic time, but differ on the details.
Of these, only Illustris and Horizon-AGN display the same overall trend of decreasing SMBH masses towards $z=0$ as we predict with \sharkII\ (EAGLE shows a declining trend only for $\sim\msun{10}$ galaxies).

The different simulations also predict different evolutions and dependency with $M_\star$ for the scatter around the relation, with some of them showing an increase in the scatter towards $z=0$ (Illustris, TNG) while others show the opposite (SIMBA) or no clear trend (EAGLE, Horizon-AGN), and with either monotonically-decreasing/increasing scatter with stellar mass (SIMBA/Illustris, Horizon-AGN) or a maximum in the scatter at $\sim\msun{10}$ (EAGLE, TNG).
The scatter in \sharkII\ shows a similar trend to that predicted by EAGLE and TNG, with a clear maximum at $M_\star\sim\msun{10}$ by $z=0$, though the location of this peak sees a strong evolution in redshift in \sharkII, being located at $M_\star\sim\msun{8}$ at $z=9$.
Another difference is that \sharkII\ predicts a second maximum in the scatter above $M_\star\gtrsim\msun{11}$, which appears by $z\sim3$, though we note that the comparatively limited volume in these hydrodynamic simulations strongly limits the number of galaxies in this mass regime.

While to the authors' knowledge there is no similar work comparing the evolution of the SMBH scaling relations across multiple SAMs, \citet{lagos2025} showed that \sharkII, \textsc{GALFORM} \citep{lacey2016}, and \textsc{GAEA} \citep{delucia2024} predict varying degrees of dependence of the scatter of the $M_\star$-$\MBH$ relation on sSFR at $z=3$.
\textsc{GALFORM} predicts a narrow relation with no difference between quenched and star-forming galaxies, \textsc{GAEA} predicts the relation breaking into two distinct ones with star-forming galaxies having smaller SMBH by more than one order of magnitude difference between the BHs of the two populations, and \sharkII\ falls in between with distinct but much closer relations ($\sim0.3$ dex difference in SMBH masses between quenched and star-forming galaxies).
We also note that \citet{porras2025} has recently compared the evolution of the BHMF from $z=0$ to $z=6$ for three different SAMs: \textsc{darksage} \citep{stevens2016}, \citet{somerville2015b}, and \citet{ricarte2018}).
While all three show some qualitative similarities between them and to \sharkII, like the progressive flattening of the high-mass end towards $z=0$, there are clear differences between the three and \sharkII\ even at $z=0$, like the number density of SMBHs with $\MBH\gtrsim\msun{9}$.

\section{Conclusions}\label{S5:summary}

In this work, we have explored the evolution of the $\Mbulge$-$\MBH$ and $M_\star$-$\MBH$ scaling relations as predicted by the \sharkII\ semi-analytic model, first presented in \citet{lagos2024} and with the updates presented in \citet{chandro2025}.
To this end, we have introduced the calibration of the model to the large-volume Planck-Millennium DM-only simulation \citep{baugh2019}, enabling the exploration of these scaling relations with extremely robust statistics.
Outside of the $z=0$ $\Mbulge$-$\MBH$ relation, which we use a secondary calibrator, we find overall good agreement of \sharkII\ with observational measurements of both relations across a wide range of redshift and stellar masses.
We predict a significant evolution in both relations as a function of cosmic time, with SMBH masses $\sim1$ dex lower at $z=0$ compared to $z=9$ at fixed stellar mass.

The scatter around these relations also evolve with time, increasing by a factor of $\sim2-5$ from $z=9$ to $z=0$, with the $\Mbulge$-$\MBH$ relation roughly exhibiting less scatter for larger stellar masses and the $M_\star$-$\MBH$ showing two distinct peaks in the scatter.
We find that the scatter in both relations show correlations with other intrinsic galaxy properties (morphology, star formation, and main source of SMBH growth), all evolving with cosmic time, suggesting different evolutionary paths as galaxies move across these relations.
In particular, the correlation with morphology predicted by \sharkII\ is in good qualitative agreement with the (comparatively more limited) observational measurements in the literature, and how a SMBH builds its mass plays a significant role in shaping these scaling relations. 
The scatter also shows a weaker correlation with galaxy environment, with low-mass satellites exhibiting systematically larger SMBH masses than central galaxies of similar mass.
Using random forest regression, we find that the differences in morphology account for most of the scatter around both scaling relations, indicating that the correlations of the SMBH relations with other galaxy properties are driven by the correlations between said properties with morphology.

\begin{acknowledgments}
MB is funded by McMaster University through the William and Caroline Herschel Fellowship.
This work was supported by resources provided by The Pawsey Supercomputing Centre with funding from the Australian Government and the Government of Western Australia.
CL thanks the ARC for the Discovery Project DP210101945.
KP acknowledges support from the Australian Government Research Training Program Scholarship.
ACG acknowledges Research Training Program and ICRAR scholarships.
This work was supported by resources provided by the Pawsey Supercomputing Research Centre’s Setonix Supercomputer (\url{https://doi.org/10.48569/18sb-8s43}) and Acacia Object Storage (\url{https://doi.org/10.48569/nfe9-a426}), with funding from the Australian Government and the Government of Western Australia.
\end{acknowledgments}

%



\software{\textsc{python} v3.11 (\url{https://www.python.org}),
          \textsc{astropy} v7.1 \citep{astropyI,astropyII,astropyIII},
          \textsc{h5py} v3.14 \citep{h5py},
          \textsc{jupyter} v1.0 \citep{jupyter},
          \textsc{matplotlib} v3.10 \citep{matplotlib},
          \textsc{numpy} v2.3 \citep{numpy},
          \textsc{pandas} v2.3 \citep{pandas},
          \textsc{scicm} v1.0 \citep{scicm},
          \textsc{scikit-learn} v1.7 \citep{scikit-learn},
          \textsc{scipy} v1.16 \citep{scipy},
          \textsc{splotch} v0.6 \citep{splotch}}

\appendix

\section{The evolution and mass dependence of the scatter around the $M_\star$-$\MBH$ and $\Mbulge$-$\MBH$ relations}\label{A1:MBH_scatter}

Here we explore the scatter around the SMBH scaling relations as predicted by \sharkII, both as a function of redshift and stellar mass, shown in Figure \ref{fig:Mstar_DMBH}.
Both relations ($\Mbulge$-$\MBH$ and $M_\star$-$\MBH$) show the same overall trend of increasing scatter towards $z=0$, with clear but distinct dependencies on stellar mass.
The evolution of the scatter in the $\Mbulge$-$\MBH$ relation show two distinct behaviours above and below $\Mbulge\sim\msun{9.5}$.
Below this mass, the scatter increases by a factor of $\sim5$ from $z=9$ ($\sim0.2$ dex) to $z=0$ ($\sim1.0$ dex).
More massive bulges start from a similar scatter ($\sim0.2$ dex) at $z=9$ but the increase in scatter stops by $z\sim1.5$ ($\sim0.4$ dex), increasing by a factor of $\sim2$ with no further evolution towards $z=0$.
We note that we find a very similar relation between $\fmerge$ and $\Mbulge$ as the one for $\fmerge$ and $\MBH$ shown in Figure \ref{fig:MBH_fmerge}, with the minimum in $\fmerge$ spanning the $10^{9.5}\lesssim\Mbulge/M_\odot\lesssim10^{10.5}$ range, suggesting that the transition seen in the scatter of the $\Mbulge$-$\MBH$ relation in Figure \ref{fig:Mstar_DMBH} could be driven by the change in how SMBH grow.

\begin{figure}
    \centering
    \includegraphics[width=\linewidth]{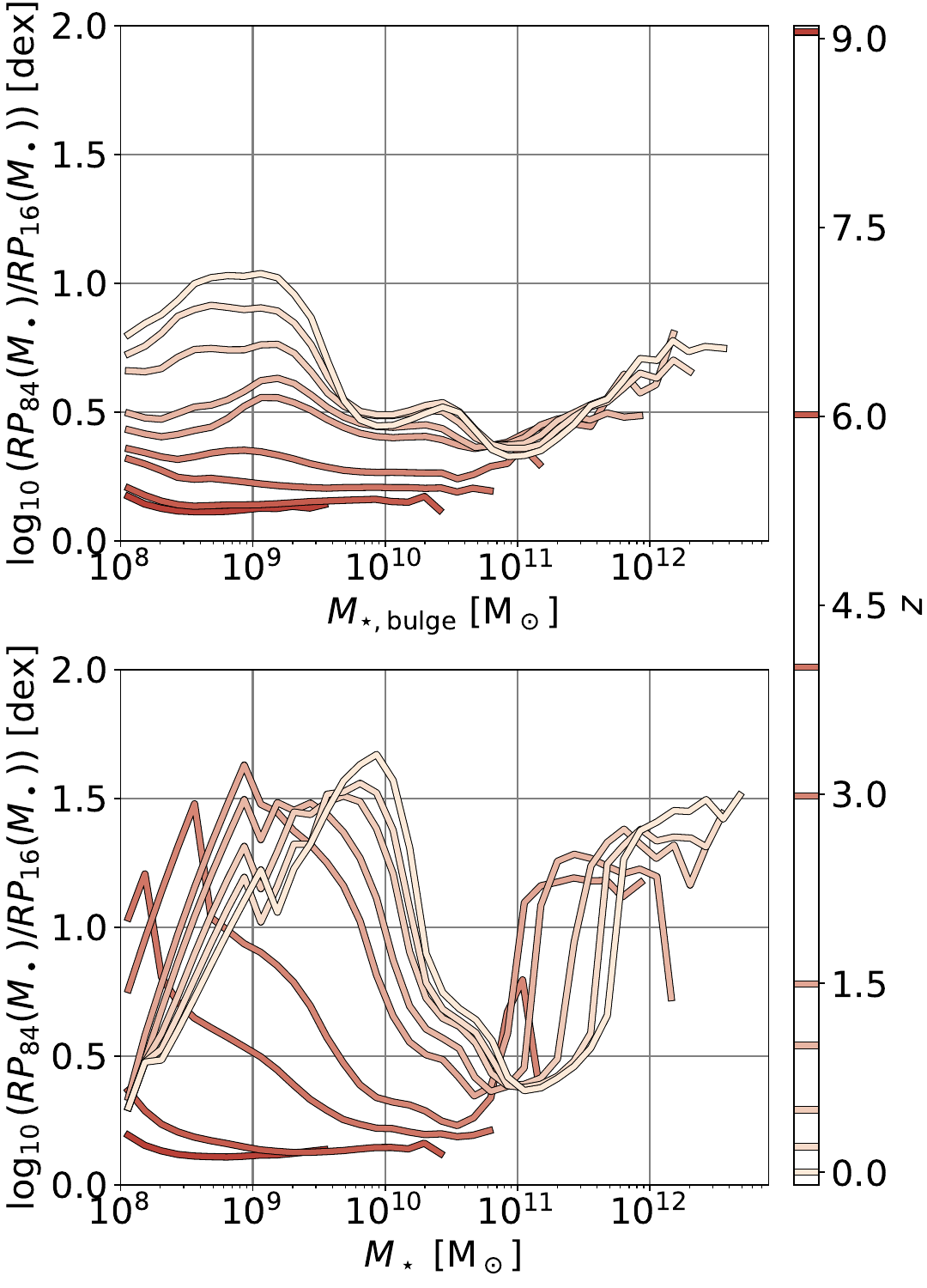}
    \caption{The scatter around the $\Mbulge$-$\MBH$ and $M_\star$-$\MBH$ relations (left and right panels, respectively), as a function of redshift and stellar mass.
    The scatter is measured as the difference in log-scale of the running 16$^\mathrm{th}$ and 84$^\mathrm{th}$ running percentiles.
    Each line indicates a different redshift, coloured as in Figure \ref{fig:Mstar_fMBH}.}
    \label{fig:Mstar_DMBH}
\end{figure}

\begin{figure*}
    \centering
    \includegraphics[width=\linewidth]{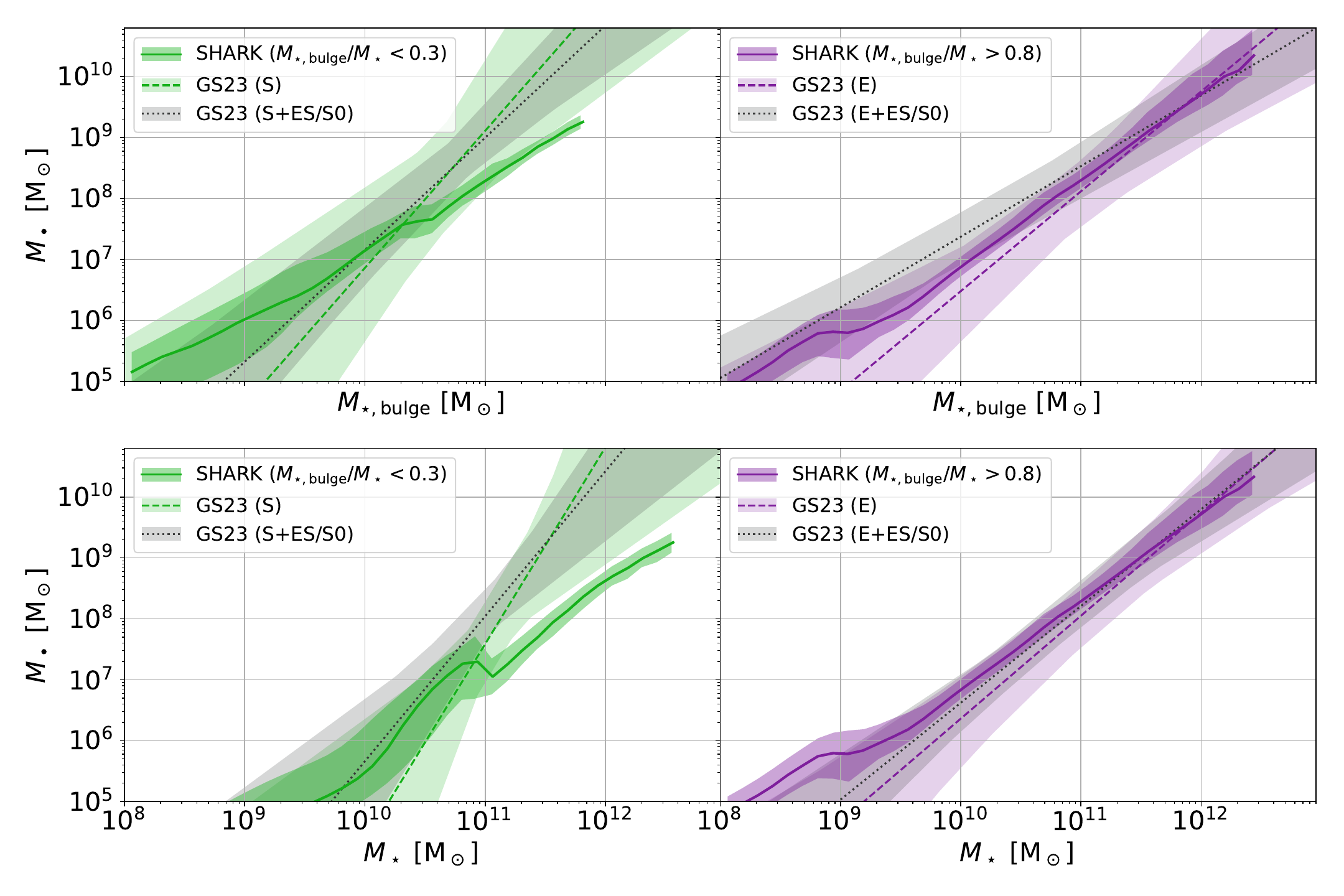}
    \caption{Comparison of the $\Mbulge$-$\MBH$ (top row) and $M_\star$-$\MBH$ (bottom row) relations predicted by \sharkII\ at $z=0$ to those measured by \citet{graham2023a}, including the \citet{chabrier2003} IMF correction from \citet{graham2024}.
    The solid lines indicated the running median in \sharkII\, with the shaded areas indicating the 16\nth-84\nth percentile range.
    The dashed and dotted lines indicated the linear fits to the relations from  \citet{graham2023a}, and the corresponding shaded areas indicated the errors in the fit.
    The left column shows the relations for late-type galaxies, and the right column for early-type galaxies.
    The dashed lines indicate the fits from \citet{graham2023a} that only include each morphological type, the dotted lines indicate those that also include ES/S0 galaxies \citep[for the definition of ES and S0, see][]{graham2023a}.}
    \label{fig:Mstar_MBH_BT_z0_plot}
\end{figure*}

In comparison, the evolution of the scatter in the $M_\star$-$\MBH$ is stronger and more complex, with two clear peaks forming by $z=0$.
The lower peak in scatter starts at $M_\star\sim\msun{8}$ at $z=9$ and moves to $M_\star\sim\msun{10}$ by $z=0$, with the 16-84\nth percentile range increasing from $\sim1.2$ dex to $\sim1.7$ dex.
The second peak appears by $z\sim3$ at $M_\star\sim\msun{11}$ and moves to $M_\star\sim\msun{12}$ at $z=0$, increasing from $\sim0.6$ dex to $\sim1.4$ dex.
The minimum between both below $z\sim3$ remains fairly stable at $\sim0.4$ dex and roughly centred at $M_\star\sim\msun{11}$, with slightly less scatter and at a lower mass at $z=3$ ($\sim0.3$ dex at $M_\star\sim\msun{10.5}$).
Unlike the case of $\Mbulge$, the mass of minimum scatter we see in $M_\star$-$\MBH$ does not seem strongly connected to $\fmerge$, as the $M_\star$ where $\fmerge$ reaches a minimum does not evolve with redshift, while the minimum in the scatter clearly evolves with redshift.

\section{Further comparison of the morphology dependence in \sharkII\ and observations}\label{A2:GS23_comp}

\begin{figure*}
    \centering
    \includegraphics[width=\linewidth]{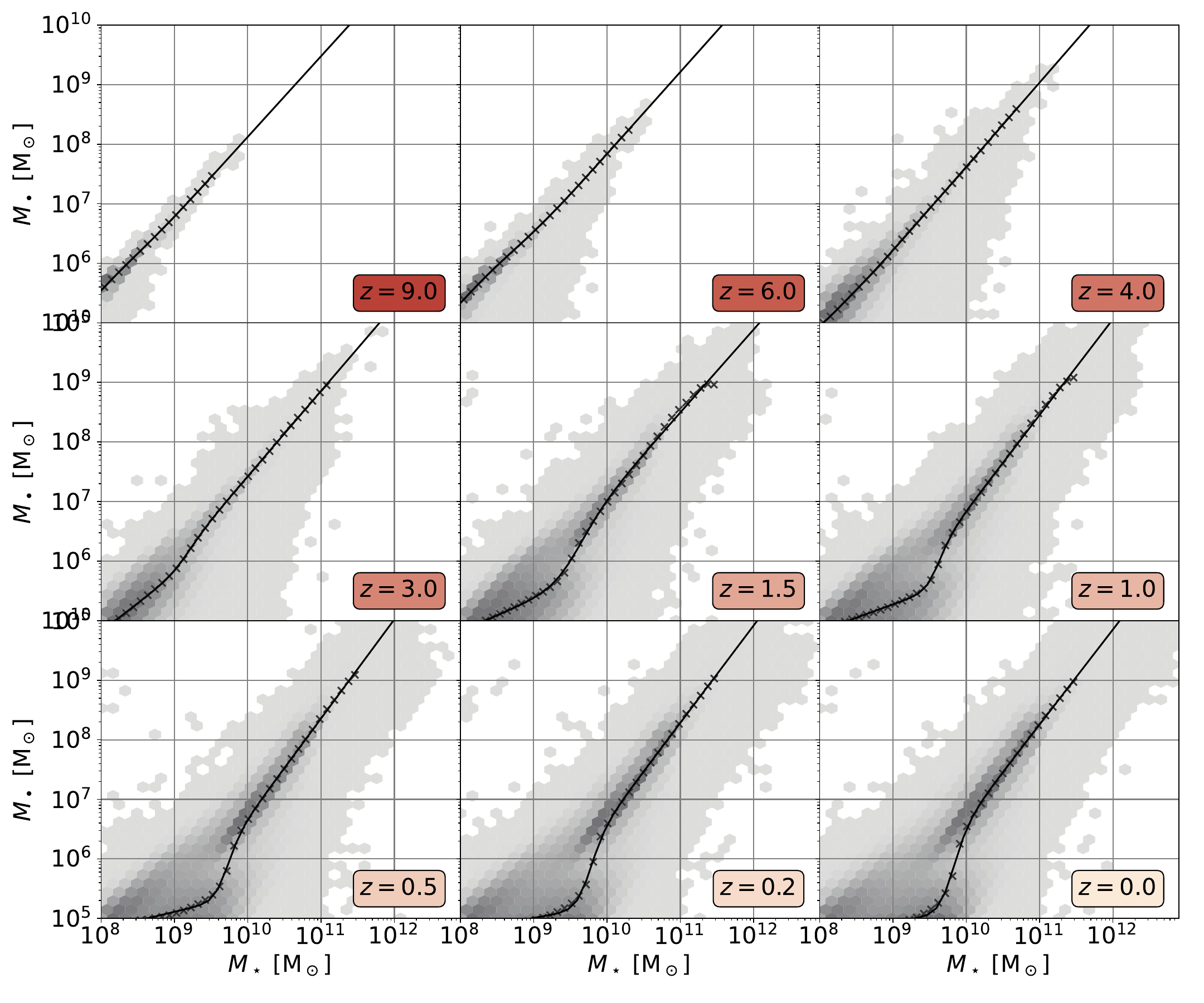}
    \caption{Fit to the $M_\star$-$\MBH$ relation used to measure $\DeltaSBHR$ used in Figure \ref{fig:FeatureImportance_MB}.
    Each panel show the $M_\star$-$\MBH$ relation in \sharkII\ as a function of redshift, with the coloured histograms in the background show the overall distribution of \sharkII\ galaxies at each redshift, cross markers showing the running median, and the solid line the fit to the relation.}
    \label{fig:MBHR_fit}
\end{figure*}

In Sections \ref{S3.1:Mbulge_MBH} and \ref{S3.2:Mstar_MBH}, we included comparisons between the trends with morphology in the $\Mbulge$-$\MBH$ and $M_\star$-$\MBH$ relations at $z=0$ in \sharkII\ to the individual local galaxy measurements from \citet{graham2023a}, but they also provide a variety of linear fits to these relations.
While we do not attempt to fully replicate their fitting procedure here, there is still value on qualitatively comparing their fits to the running medians we measure from \sharkII, which we show in Figure \ref{fig:Mstar_MBH_BT_z0_plot}.
The relation for early-type galaxies shows a good agreement between the observational results of \citet{graham2023a} and our predictions from \sharkII, with the running medians and 16\nth-84\nth percentiles comfortably within the errors reported by \citet{graham2023a}\footnote{It is intriguing that the $\Mbulge$-$\MBH$ relation in \sharkII\ is in better agreement with the more pure E fit from \citet{graham2023a}, while for the $M_\star$-$\MBH$ relation the agreement is better with the more contaminated E+ES/S0 fit, but we do not attempt to extend the discussion of this beyond noting this discrepancy.}.
We find a similar agreement for late-type galaxies, but only for stellar masses below $\sim\msun{10.5}$ (for both $M_\star$ and $\Mbulge$), above which \sharkII\ predicts galaxies below the linear fits from \citet{graham2023a}.
We note though that the limited number of S/ES/SO galaxies with $\Mbulge\gtrsim\msun{10.5}$ in \citet{graham2023a} all lie below their fits (see their figure 3), which is qualitatively consistent with the predictions from \sharkII, although the same is not true when considering $M_\star$ instead (suggesting a real tension with observations).

\section{Defining the offset from the different mass scaling relations}\label{A3:mass_relations}

In this appendix, we describe how we measure the offset from the different relations discussed and presented in this work: the SFMS (Section \ref{S3.2:Mstar_MBH} and Figure \ref{fig:Mstar_MBH_SFMS}), SMSR (Section \ref{S3.2:Mstar_MBH})
, and $\Mbulge$-$\MBH$ and $M_\star$-$\MBH$ relations (Section \ref{S4:disc} and Figure \ref{fig:FeatureImportance_MB}).
First, we measure the running medians of each property ($r^{}_{50,\star}$, $\dot{M}_\star/M_\star$, $\Mhalo$, and $\MBH$) as a function of $\Mbulge$ or $M_\star$ at every snapshot, across 50 equally-sized mass bins in the range $10^8<M_\star/\mathrm{M}_\odot<10^{12.9}$.
For the SFMS, we apply two additional selections to ensure well-behaved running medians and avoid contamination from quiescent galaxies, reducing the mass range to a maximum of $\msun{11}$ and only selecting galaxies with a sSFR greater than $10^{-11}$ yr$^{-1}$. 

We fit the resulting running median with the following function \citep[inspired by equations 10 and 11 in][]{taylor2015}:
\begin{equation}
    f(x) = (ax+b)+\tanh \left(\frac{x-x_0}{s}\right)(cx+d).
\end{equation}
\noindent We note that this equation converges to $f(x) = (a-c)x+(b-d)$ and to $f(x) = (a+c)x+(b+d)$ when $x$ goes to positive and negative infinity, respectively, with the transition point being $x_0$, and $s$ controlling how fast/smooth the transition is (larger values of $s$ produce a more gradual transition).
In practice, this equation behaves similar to a broken power law fit to our data, with the main benefit being additional control on the smoothness of the transition between slopes.
While not all relations require this greater flexibility, this equation does provide a better fit to some relations that exhibit a more complex behaviour, as exemplified by the $M_\star$-$\MBH$ relation in Figure \ref{fig:MBHR_fit}).
We use these fits to measure the vertical offsets from these relations for all galaxies as: 
\begin{align}
    \DeltaSFMS &=\log_{10}\left( \dot{M_{\star}} / f^{}_\mathrm{SFMS}(M_{\star}) \right), \\
    \DeltaMSR &=\log_{10}\left( r_{50,\star} / f^{}_\mathrm{SMSR}(M_{\star}) \right), \\
    \DeltaSBHR &=\log_{10}\left( \MBH / f^{}_{\Mbulge-\MBH}(M_{\star}) \right), \\
    \DeltaBBHR &=\log_{10}\left( \MBH / f^{}_{M_\star-\MBH}(M_{\star}) \right),
\end{align}
\noindent where $\dot{M_{\star}}$ is the galaxy SFR.
Figure \ref{fig:MBHR_fit} shows the \sharkII\ $M_\star$-$\MBH$ relation and the corresponding fits as a function of redshift.

\bibliography{papers}{}
\bibliographystyle{aasjournal}


\end{document}